\newif\ifdraft\draftfalse 
\newif\ifanon\anonfalse    
\newif\ifreview\reviewfalse

\draftfalse

\documentclass[a4paper,UKenglish,cleveref, autoref, thm-restate]{lipics-v2021}

\pdfoutput=1 
\hideLIPIcs  

\title{Nominal Type Theory by Nullary Internal Parametricity} 

\author{Antoine {Van Muylder}}{DistriNet, KU Leuven, Belgium}{antoine.vanmuylder@kuleuven.be}{https://orcid.org/0000-0003-4144-9368}{}
\author{Andreas Nuyts}{DistriNet, KU Leuven, Belgium}{andreas.nuyts@kuleuven.be}{https://orcid.org/0000-0002-1571-5063}{}
\author{Dominique Devriese}{DistriNet, KU Leuven, Belgium}{dominique.devriese@kuleuven.be}{https://orcid.org/0000-0002-3862-6856}{}

\authorrunning{A. Van Muylder, A. Nuyts, and D. Devriese}

\Copyright{Antoine Van Muylder, Andreas Nuyts, Dominique Devriese} 



\funding{Andreas Nuyts holds a Postdoctoral Fellowship from the Research Foundation - Flanders (FWO; 12AB225N). This research is partially funded by the Research Fund KU Leuven, the Internal Funds KU Leuven and the FWO (G027225N).}

\acknowledgements{We thank the reviewers for their time and helpful comments.}

\nolinenumbers 

\usepackage{inconsolata} 

\lstdefinestyle{mystyle}{
    basicstyle=\fontfamily{zi4}\selectfont\small, 
    numbers = none,
    breakatwhitespace=false,         
    breaklines=true,                 
    captionpos=b,                    
    keepspaces=true,                 
    numbersep=5pt,                  
    showspaces=false,                
    showstringspaces=false,
    showtabs=false,                  
    tabsize=2
}
\usepackage{stmaryrd}

\usepackage{bm} 

\usepackage{bussproofs}

\usepackage{xcolor}
\usepackage{soul}
\colorlet{mathgray}{gray!25}

\allowdisplaybreaks

\crefformat{footnote}{#2\footnotemark[#1]#3}

\makeatletter
\newcommand{\IE}{i.e.\@}
\newcommand{\EG}{e.g.\@}

\makeatother

\newcommand{\op}{^{\mathsf{op}}}

\newcommand{\rmSet}{{\mathrm{Set}}}

\newcommand\aint{@\mathsf{N}}
\newcommand\inti{\mathsf{N}}
\newcommand\nm{\mathsf{Nm}}
\newcommand\varinctx[2]{(#2 : \aint)\in #1} 
\newcommand\indN{\mathsf{ind}_\nm}

\newcommand{\namesym}{\text{\reflectbox{\rotatebox[origin=c]{180}{$\mathsf{N}$}}}}

\newcommand{\sysSchopp}{BNTT}
\newcommand{\sysCheney}{$\lambda^{\Pi\namesym}$}

\newcommand{\angles}[1]{\mathopen{}\left\langle #1 \right\rangle\mathclose{}}

\newcommand{\accol}[1]{\mathopen{}\left\{ #1 \right\}\mathclose{}}
\newcommand{\paren}[1]{\mathopen{}\left( #1 \right)\mathclose{}}

\newcommand\rlab[1]{\RightLabel{\scriptsize{\textsc{#1}}}}
\newcommand\sftype{\mathsf{type}}

\newcommand\cU{\mathcal{U}}
\newcommand\Ltm{\mathsf{Ltm}}

\newcommand\cart{\mathsf{c}}
\newcommand\restr{\textbackslash}
\newcommand\sfext{\mathsf{ext}}
\newcommand\sfGel{\mathsf{Gel}}
\newcommand\sfgel{\mathsf{gel}}
\newcommand\sfung{\mathsf{ung}}

\newcommand\sfLtm{\mathsf{Ltm}}
\newcommand\sfvar{\mathsf{var}}
\newcommand\sfapp{\mathsf{app}}

\newcommand\sfbind{\mathsf{bind}}
\newcommand\sfswap{\mathsf{swap}}

\newcommand\sfubd{\mathsf{ubd}}
\newcommand\sftoh{\mathsf{toh}}
\newcommand\Henc{\mathsf{HEnc}}
\newcommand\HMod{\mathsf{HMod}}
\newcommand\NMod{\mathsf{NMod}}
\newcommand\mknm{\mathsf{mkNM}}
\newcommand\mkhm{\mathsf{mkHM}}

\newcommand\sflb{\mathsf{lbump}}
\newcommand\sfmb{\mathsf{mbump}}
\newcommand\sfeb{\mathsf{ebump}}
\newcommand\sfnb{\mathsf{nbump}}

\definecolor{dkblue}{rgb}{0,0.1,0.5}
\definecolor{dkgreen}{rgb}{0,0.4,0}
\definecolor{dkred}{rgb}{0.6,0,0}
\definecolor{lred}{rgb}{0.70,0,0}
\definecolor{dkpurple}{rgb}{0.7,0,1.0}
\definecolor{purple}{rgb}{0.9,0,1.0}
\definecolor{olive}{rgb}{0.4, 0.4, 0.0}
\definecolor{teal}{rgb}{0.0,0.4,0.4}
\definecolor{azure}{rgb}{0.0, 0.5, 1.0}
\definecolor{btbtgray}{rgb}{0.9, 0.9, 0.9}
\definecolor{btgray}{rgb}{0.8, 0.8, 0.8}
\definecolor{gray}{rgb}{0.5, 0.5, 0.5}
\definecolor{dkgray}{rgb}{0.3, 0.3, 0.3}
\definecolor{agdablue}{RGB}{0, 0, 205}
\definecolor{agdapink}{RGB}{205, 0, 205} 

\begin{document}

\maketitle

\begin{abstract}
There are many ways to represent the syntax of a language with binders. In particular, nominal frameworks are metalanguages that feature (among others) name abstraction types, which can be used to specify the type of binders. The resulting syntax representation, nominal data types, makes alpha-equivalent terms equal, has a closed induction principle and is well-behaved w.r.t.\ weakening. It is known that name abstraction types can be presented either as an existential or as a universal quantification on names. Existential name abstractions support matching on name-binding patterns but have cumbersome typing rules; universal ones have clean rules but apparently no such nominal matching. In this work we show that this matching ability and other nominal features are recovered in a type theory consisting of (1) Nullary ($0$-ary) Internally Parametric Type Theory (nullary PTT), (2) a type of names and a novel name induction principle, (3) nominal data types. This type theory is a legitimate nominal framework: it has universal and (non-primitive) existential name abstractions, a freshness type former, restricted name swapping and local-scope operations. Nominal pattern matching is recovered via term-relevant nullary parametricity. We provide an example involving synthetic Kripke parametricity.
\end{abstract}




\section{Introduction}\label{sec:intro}

There are many ways to formally define the syntax of a language with binders.
In particular,
nominal frameworks~\cite{DBLP:journals/iandc/Pitts03, DBLP:conf/icfp/ShinwellPG03, DBLP:journals/entcs/PittsMD15, schopp2004dependent,
DBLP:journals/corr/abs-1201-5240, DBLP:journals/jar/Urban08}
are metalanguages featuring (among others) a primitive type of names $\nm$ as well as a ``name abstraction'' type former
which we write $\aint \multimap {-}$.
Name abstraction types can be used to specify the type of binders of a given object language.
For example, we can define the syntax of the untyped lambda calculus (ULC) with the following data type.
\begin{lstlisting}[mathescape]
data Ltm : $\cU$ where
  var : $\nm$ $\to$ Ltm
  app : Ltm $\to$ Ltm $\to$ Ltm
  lam : ($\aint$ $\multimap$ Ltm) $\to$ Ltm
\end{lstlisting}

The resulting representation is called a ``nominal data type'' or ``nominal syntax''
and offers several advantages.
First, $\alpha$-equivalent terms are definitionally equal, a consequence
of how $\aint \multimap {-}$ is axiomatized.
Second, nominal data types have a (closed) induction principle.
Third, by contrast with the De Bruijn representation,
weakening by a name simply means extending the context with that name,
and all programs in the metalanguage are automatically
stable under this operation.
Fourth, formal reasoning with nominal syntax can sometimes
match on-paper reasoning~\cite{DBLP:journals/jar/Urban08}.
A drawback of this representation is that it is not definable in plain type theory.

Inference rules for the name abstraction type former can be defined in two ways:
either in an existential/positive, or in a universal/negative fashion.
Interestingly, these presentations are semantically
equivalent~\cite{schopp2004dependent, DBLP:phd/ethos/Schopp06} but syntactically have different pros and cons.

On the one hand, the existential presentation makes the name abstraction type former
behave as an existential quantification on names.
So an element of that type, i.e.\ a name abstraction,
can be understood as a \emph{name-term pair up to $\alpha$-renaming}  $\langle x, a \rangle$ where
$a$ may bind the freshly chosen name $x$. 
This presentation is convenient since
the user is allowed to \emph{pattern match}
on such ``binding'' name-term pairs, an ability that we call nominal pattern matching.
The following example is Example 2.1 from~\cite{DBLP:conf/icfp/ShinwellPG03}
and illustrates this ability. It is a (pseudo-) FreshML program computing the equality
modulo $\alpha$-renaming of two existential name abstractions. In the example
$A$ has decidable equality $\mathsf{eq}_A : A \to A \to \mathsf{Bool}$ and
the existential name abstraction type is written $\aint \cdot {-}$.
\begin{lstlisting}[mathescape]
eqabs : ($\aint$ $\cdot$ $A$) $\to$ ($\aint$ $\cdot$ $A$) $\to$ $\mathsf{Bool}$
eqabs $\langle x_0,a_0\rangle$ $\langle x_1,a_1\rangle$ = $\mathsf{eq}_A$ (swap $x_0,x_1$ in $a_0$) $a_1$
\end{lstlisting}
Note (1) the occurrences of $x_0,x_1$ outside of $\langle x_0,a_0\rangle,\langle x_1,a_1\rangle$ and
(2) the appearance of the $\mathsf{swap}$ operation exchanging free names.

Nominal pattern matching is convenient and for that reason
nominal frameworks use the existential presentation
for practical reasoning about nominal syntax.
However,
inference rules for existential abstraction types are cumbersome
to specify. The issue is that a name $x$ is considered
fresh in a binding pair $\langle x,a\rangle$, and such information must be
encoded at the type level and propagated when pattern matching.
The consequence is that the rules for existential abstraction types are polluted with
typal freshness information.
This makes the implementation of such rules harder in a proof assistant environment.

On the other hand,
the universal presentation treats name abstractions not as pairs, but
rather as \emph{functions} consuming fresh names (a.k.a. affine or fresh functions),
as in~\cite{DBLP:journals/entcs/PittsMD15, DBLP:journals/corr/abs-1201-5240}.
This has several consequences:
names can
only be used when they are in scope,
no explicit swapping primitive is needed
and the inference rules are straightforward to specify.
However one seems to lose the important ability to pattern match.

In this paper we will propose a new foundation for nominal frameworks
alleviating the above issues and relying
on the notion of \emph{parametricity}.
Concisely put, $n$-ary parametricity asserts that types behave as $n$-ary reflexive graphs.
More precisely, the $n$-ary parametricity of a type $T$ computes
by induction on $T$ a certain
$n$-ary reflexive graph structure for $T$.
Even more precisely,
(1) the $n$-ary parametricity of a (open) type $T$ computes to a certain (open)
$n$-ary relation written $[T]_n : T\to\ldots\to T \to \cU$,
(2) the $n$-ary parametricity of a (open) term $t:T$ computes to a certain reflexivity edge at $t$,
\IE{} $[t]_n : [T]_n \,t\,\ldots\,t$.
Parametricity is most often used in its binary form, at types of polymorphic
functions, where it specializes to a uniformity
property~\cite{reynolds1983types, DBLP:conf/fpca/Wadler89}.
For example if $t : T = \forall (X: \cU).\,X \to X$ then
$[t]_2 : [T]_2 \,t \,t$ is a proof that $t$ maps related types to related functions.
Indeed $[-]_2$ is defined by induction (not given here) and $[T]_2\,t\,t$ reduces to
$\allowbreak\forall (X_0 X_1 : \cU)\allowbreak (R : X_0 \to X_1 \to \cU).\allowbreak\forall x_0\,x_1.\allowbreak\, R\, x_0 \, x_1 \to\allowbreak R\,(t \,X_0 \,x_0)\,\allowbreak(t\,X_1\,x_1)$.
Intuitively this property disallows $t$ from inspecting its type argument $X:\cU$.
It can be shown that the property entails that $t$ equals the identity $\lambda \,X\,x.\,x$.

%

In the above account, parametricity is regarded as a property that is defined
and proven to hold \emph{externally} about a dependent type theory (DTT), as in~\cite{bernardy2012proofs}.
In other words,
the parametricity property, or translation,
is a map $[-]_n$ defined by induction on the (open) types and terms of DTT.
By contrast, $n$-ary \emph{internally} parametric type theory
($n$-ary PTT for short)~\cite{DBLP:journals/lmcs/CavalloH21, moulin2016internalizing, nuyts_degrees_2018, nuyts2017parametric,
DBLP:journals/pacmpl/AltenkirchCKS24, DBLP:journals/corr/abs-2404-03825}
extends DTT with new type and term formers. These primitives make
it possible to prove parametricity results \emph{within} $n$-ary PTT.

To that end, $n$-ary PTT typically provides two kinds of primitives
internalizing aspects of reflexive graphs.
Firstly, Bridge types are provided, which intuitively are to a type what an edge is to a reflexive graph.
Syntactically, a bridge $q : \mathsf{Bridge}\,A\,a_0\,\ldots\,a_{n-1}$
at type $A$ between $a_0,\ldots,a_{n-1}$
is treated as a function $\inti \to A$ out of a posited bridge
interval $\inti$. The $\inti$ interval contains $n$ endpoints $(e_i)_{i<n}$ and $q$
must respect these definitionally $q\,e_i = a_i$.
When $n = 2$ this is similar to the ``paths'' of cubical type theory~\cite{vezzosi_cubical_2019}
which play the role of equality proofs in the latter.
However, bridges do not satisfy various properties of paths, e.g.\ they cannot
be composed and one cannot transport values of a type $P~x$ over a bridge $\mathsf{Bridge}~A~x~y$ to type $P~y$.
Moreover, contrary to paths, bridges may only be applied to variables $x: \inti$ that
do not appear freely in them, i.e.\ that are fresh for the bridge.
Functions with such a freshness side condition are also known as affine or fresh functions and we will use the symbol $\multimap$ instead of $\to$ for their type.
Secondly, the other primitives of n-ary PTT make it possible to
prove that the Bridge type former has a commutation law with respect to every other type former.
\cref{fig:commutation-table} lists some of these laws (equivalences)
for arity $n = 2$ and compares the situation for bridges and paths. Path types are written $\equiv$.

The last column in \cref{fig:commutation-table} excerpts
a collection of equivalences known as the Structure
Identity Principle (SIP) in HoTT/UF~\cite{DBLP:books/daglib/0046165}. Concisely, it states
that at every type $K$,
equality is equivalent to observational equality~\cite{DBLP:conf/plpv/AltenkirchMS07}.
The Bridge column of \cref{fig:commutation-table} shows an analogous Structure Relatedness Principle (SRP)~\cite{DBLP:journals/pacmpl/MuylderND24} which
expresses equivalence of $K$'s Bridge type to the parametricity
translation of $K$, i.e.\ the Bridge type former internalizes the parametricity translation for types \emph{up to SRP equivalences}.
Accordingly, in this setting the parametricity of a term $k$ is $k$'s reflexivity bridge $\lambda (\_:\inti).\,k : \mathsf{Bridge}\,K\,k\,\ldots\,k$
mapped through the SRP equivalence at $K$.

\begin{figure}
\begin{centering}%
\begin{tabular}{ | c || p{6.7cm} | p{5cm} |}
\hline
$K : \cU$  & $\mathsf{Bridge}\,K\,k_0\,k_1 \simeq\ldots$ & $k_0 \equiv_K k_1 \simeq \ldots$\\
\hline\hline
$A \to B$ & $\forall a_0\,a_1\ldotp\,\mathsf{Bridge}\,A\,a_0\, a_1\to \mathsf{Bridge}\,B\,(k_0\,a_0)\,(k_1\,a_1)$ & $\forall\,a_0\,a_1\ldotp a_0 \equiv_A a_1 \to k_0\,a_0 \equiv_Bk_1\,a_1$\\
\hline
$A \times B$ & $\mathsf{Bridge}\,A\,(k_0 \,\mathsf{.fst}) (k_1 \,\mathsf{.fst}) \times \newline \mathsf{Bridge}\,B\,(k_0 \,\mathsf{.snd}) (k_1 \,\mathsf{.snd})$ &$(k_0 \,\mathsf{.fst}) \equiv_A (k_1 \,\mathsf{.fst}) \times \newline (k_0 \,\mathsf{.snd}) \equiv_B (k_1 \,\mathsf{.snd})$\\
\hline
$\cU$ & $k_0 \to k_1 \to \cU$ & $k_0 \simeq k_1$\\
\hline
\end{tabular}
\caption{The Bridge and Path type formers commute with some example type formers.}
\label{fig:commutation-table}
\end{centering}
\end{figure}

\subparagraph*{Parametric Nominal Type Theory}
In this work, we put forward and elaborate an idea that existed in the community~\cite{narya_nominal2025, DBLP:journals/lmcs/NuytsD24}, which is to use
nullary PTT (i.e.\ 0-ary PTT) as the foundation for nominal dependent type theory.
Specifically, we contribute Parametric Nominal Type Theory (PNTT).
Concisely put, PNTT = nullary PTT + a type of names with an induction principle + rules for nominal data types.

In more detail, PNTT is obtained from nullary PTT in several steps.
First, we make the simple observation that universal name abstraction types have the same rules
as the nullary Bridge types from nullary PTT: nullary bridges and universal abstractions are simply affine functions out of the interval.
Second, we show that, on its own, nullary PTT already supports important
features of nominal frameworks:
universal name abstractions (bridge types),
typal freshness,
(non-primitive) existential name abstractions,
a name swapping operation
and the local-scoping primitive $\nu$ of~\cite{DBLP:journals/entcs/PittsMD15}
(roughly, the latter primitive is used to witness freshness).
Third, we can recover nominal pattern matching in this parametric
setting by extending nullary PTT with a full-fledged type of names $\vdash \nm$.
It comes equipped with a novel eliminator called \emph{name induction},
which expresses for a term  $\Gamma \vdash n : \nm$ and a bridge variable $x$ in $\Gamma$,
that either $n$ is just $x$, or $x$ is \emph{fresh} in $n$.
Fourth, PNTT also has rules for the nominal data types we wish to study, like the $\Ltm$ type.

Nominal pattern matching can be indirectly recovered via an extended version of the nullary SRP:
the Structure Abstraction Principle (SAP), as we call it.
The SAP expresses that the nullary bridge type former $\aint \multimap {-}$ commutes with
every type former in a specific way, including (1) standard type formers,
(2) the $\nm$ type and (3) nominal data types.
For standard type formers, the SAP asserts that name abstraction
commutes as one might expect, e.g.\ $(\aint \multimap A \to B) \simeq ((\aint \multimap A) \to (\aint \multimap B))$.
For $\nm$, the SAP asserts that $(\aint \multimap \nm) \simeq 1 + \nm$,
which can be proved by name induction.
The SAP also holds for nominal data types.
For example, the SAP for $\Ltm$ asserts that
there is an equivalence $e : (\aint \multimap \Ltm) \simeq \Ltm_1$ between $\aint \multimap \Ltm$ and the following data type (this was proved semantically by Hofmann~\cite{DBLP:conf/lics/Hofmann99}).
Intuitively, it corresponds to $\Ltm$ terms with $\nm$-shaped holes
and we say it ought to be the nullary parametricity translation of $\Ltm$.
\begin{lstlisting}[mathescape]
data $\Ltm_1$ : $\cU$ where
  hole : $\Ltm_1$
  var : $\nm$ $\to$ $\Ltm_1$
  app : $\Ltm_1$ $\to$ $\Ltm_1$ $\to$ $\Ltm_1$
  lam : ($\aint$ $\multimap$ $\Ltm_1$) $\to$ $\Ltm_1$
\end{lstlisting}
When matching on $\mathsf{lam}\,g:\Ltm$ we have $g : \aint \multimap \Ltm$ and equivalently $e\, g : \Ltm_1$ for which the induction principle of $\Ltm_1$ applies.

\subparagraph*{Contributions and Outline}
In this paper,
we propose nullary PTT as the foundation for nominal dependent type theory, and provide
evidence for the suitability of this foundation.
\begin{itemize}
\item In Section~\ref{sec:nullary-ptt}, we
present Parametric Nominal Type Theory (PNTT).
PNTT is a variant of the univalent parametric type theory of Cavallo and Harper~\cite{DBLP:journals/lmcs/CavalloH21} where
  (1) we replace the arity $n = 2$ by $n = 0$ (see \cref{sec:nullary-ch}),
  (2) we construct a type $\nm$ from the bridge interval $\aint$ and provide a name induction principle (see \cref{sec:name-ind}),
  (3) we add rules for the specific nominal data types we wish to study.
  We explain how the above primitives validate our Structure Abstraction Principle (SAP).
  We discuss semantics and soundness in \cref{sec:semantics-soundness}.

\item In Section~\ref{sec:nomin-prim-free}, we systematically discuss the inference rules of existing nominal frameworks for typal freshness, name swapping, the local-scoping primitive $\nu$, existential and universal name abstractions.
  We explain in detail how they can all be (adapted and) implemented in terms of nullary PTT primitives.

\item Section~\ref{sec:case-studies} shows nominal techniques in PNTT in action.
\cref{sec:pi-calc-example} shows that the nullary translation of data types
lets us indirectly emulate nominal pattern matching:
defining functions by matching on patterns which bind variables.
\cref{sec:hoas-example} connects the $\Ltm$ nominal data type to a nominal HOAS representation,
and serves two purposes.
First, it illustrates that more often
than not, proving the correctness of a function $f : D \to E$ defined
by recursion out of a nominal data type $D$ requires computing the parametricity translation of $f$, \IE{}
the parametricity
of $f$ is term-relevant.
Second, the example sheds some light on the notion of
\emph{Kripke} binary parametricity~\cite{DBLP:conf/tlca/Atkey09}.
A proof in Atkey's Kripke model is ported to PNTT, and the Kripke aspect of it is emulated by nullary parametricity.
\end{itemize}
Binary parametric cubical type theory has already been implemented~\cite{DBLP:journals/pacmpl/MuylderND24}.
Thus our work offers a clear path to a practical implementation of nominal type theory.
To our knowledge, we are also the first to make explicit and active use of nullary parametricity.
We discuss related work in more detail in Section~\ref{sec:related-work}.

\section{Parametric Nominal Type Theory (PNTT)}\label{sec:nullary-ptt}
PNTT is largely based on the binary PTT of Cavallo and Harper (CH type theory, \cite{DBLP:journals/lmcs/CavalloH21}). Concretely, our rules are obtained by
(1) considering the rules of the parametricity primitives of the CH binary PTT
and replacing the arity $2$ by $0$ instead,
(2) adding novel rules to turn the bridge interval $\aint$ into a full-fledged type $\nm$, including a name induction principle and
(3) adding rules for the desired nominal data types.
For the impatient reader, the relevant rules of PNTT
appear in \cref{fig:nullary-ptt-rules} and $\mathsf{Gel}\, A\, x$
can be understood as the elements of $A$ for which $x$ is fresh. 

\subparagraph*{Cubical type theory}
Apart from its parametricity primitives, the CH theory is a cubical type theory in the way
it handles equality proofs (\emph{paths}), and so is our system.
Cubical type theory (cubical TT, \cite{cohen_cubical_2016, DBLP:journals/mscs/AngiuliBCHHL21})
is a form of homotopy type theory (HoTT, \cite{DBLP:books/daglib/0046165})
that adds new types, terms and equations on top of
plain dependent type theory. These extra primitives make it possible,
among other things, to prove univalence and more generally the SIP (see~\cref{fig:commutation-table})
at any concrete, fully known type.

The SIP matters to us because, \EG{}, univalence is used to characterize the bridge type at the universe
$(\aint \multimap \cU) \simeq \cU$.
Otherwise, the fact that our system is a cubical type theory is not of primary significance,
although we wish to remain as close as possible to the CH theory
to be able to reuse their proofs and semantics.

Accordingly, we only need to know that the primitives introduced by cubical type theory
validate the SIP in order to showcase our theorems and examples.
Yet we list these primitives for completeness:
(1) the path interval $\mathrm{I}$, dependent path types and their rules; paths play the role of equality proofs
thus non-dependent path types are written $\equiv$,
(2) the transport and cube-composition operations, a.k.a the Kan operations; these are used e.g.\ to
prove transitivity of the path relation,
(3) a type former to turn equivalences into paths in the universe, validating one direction of univalence,
(4) higher inductive types (HITs) if desired.

Additionally, in HoTT an equivalence $A \simeq B$ is by definition a function $A \to B$
with contractible fibers. We rather build equivalences using
Theorem 4.2.3 of~\cite{DBLP:books/daglib/0046165}: let $f : A\to B$ have a
quasi-inverse, i.e.\ a map $g : B \to A$ satisfying the two roundtrip equalities $\forall a.\,g(f\,a) \equiv a$ and $\forall b.\,f(g\,b)\equiv b$.
Then $f$ can be turned into an equivalence $A\simeq B$.

\subsection{Nullary CH}\label{sec:nullary-ch}

\begin{figure}[htb]
\begin{footnotesize}
\renewcommand{\defaultHypSeparation}{\hspace{0pt}}
\input{nullary-ptt-rules.tex}
\end{footnotesize}
\caption{
Parametricity primitives of PNTT.
Prem.\ of \textsc{x} - $y$ means premises of rule \textsc{x} except $y$.
For binding rules such as \textsc{ext}, we rely on the invertibility of both variable and name abstraction to bind using $\lambda$.}
\label{fig:nullary-ptt-rules}
\end{figure}

Let us explain the rules of \cref{fig:nullary-ptt-rules}. 
We begin with the nullary primitives of the CH type theory since our rules for
the bridge interval $\nm$ depend on them.

\subparagraph*{Contexts}
There are two ways to extend a context. The first way is the usual ``cartesian'' comprehension operation where $\Gamma$
gets extended with a type $\Gamma \vdash A \, \sftype$ resulting in a context $\Gamma, (a:A)$.
The second, distinct way to extend a context $\Gamma$ is an ``affine'' comprehension operation where
$\Gamma$ gets extended with a bridge variable $x$ resulting in a context written $\Gamma, (x:\aint)$.
Note that $\aint$ is \emph{not} a type and morally always appears on the left of $\vdash$ (the
Bridge type former will be written $\aint \multimap {-}$ but this is just a suggestive notation).

The presence of $\aint$ is required because
the theory treats affine variables in a special way that is ultimately
used to prove the SAP (briefly mentioned in \cref{sec:intro}).
Specifically,
terms, types and substitutions depending on an affine variable $x:\aint$
are not allowed to duplicate $x$ in affine positions.
So typechecking an expression that depends on $x:\aint$ may involve
checking that $x$ does not appear freely in some subexpressions.
Since variables declared after $x$ in the context
may eventually be substituted by terms mentioning $x$,
a similar verification must be performed for them.
More formally, such ``freshness'' statements about free variables are specified
in the inference rules
using a context restriction operation ${-}\restr{-}$ (as in~\cite{DBLP:journals/lmcs/CavalloH21}, section 2.1).
If $\Gamma$ is a context containing $(x:\aint)$ then
$\Gamma \restr x$ is the context obtained from $\Gamma$ by removing $x$ itself,
as well as all the cartesian (i.e.\ not $\aint$) variables to the
right of $x$.\footnote{A detail: variables $i:\mathrm{I}$ and cubical constraints
are not removed~\cite{DBLP:journals/lmcs/CavalloH21, DBLP:journals/pacmpl/MuylderND24} as they cannot depend on $x$.}
If $\varinctx{\Gamma}{x}$ and $\Gamma\restr x \vdash a :A$ we say that $x$ is fresh in $a$.

\subparagraph*{Nullary bridges}
The type former of dependent bridges with dependent codomain
$\Gamma,(x:\aint)\vdash A$ is written $(x:\aint)\multimap A$.
The type of non-dependent bridges with codomain $\Gamma \vdash A$
is written $\aint \multimap A$ and defined as $({\_} : \aint) \multimap A$.
Introducing a bridge requires providing a term in a context extended with
a bridge variable $(x:\aint)$.
A bridge $a' : (y:\aint) \multimap A$ can be eliminated at a variable $(x:\aint)$ only
if $x$ is fresh in $a'$.
In summary, nullary bridges are treated and written like (dependent)
functions out of the $\aint$ pretype
but are restricted to consume fresh variables only.
From a nominal point of view, a bridge $a' : \aint \multimap A$ is a name-abstracted value in $A$.

Since we will use the theory on non-trivial examples in the next sections,
we prefer to explain the other primitives of \cref{fig:nullary-ptt-rules}
from a user perspective:
we derive programs corresponding to
these primitives 
in the empty context  and
express types as elements of the universe type $\cU$, whose rules are not listed but standard.
Furthermore we explain what the equations of \cref{fig:nullary-ptt-rules}
entail for these closed programs.
The closed programs derived from the CH nullary primitives
have a binary counterpart in~\cite{DBLP:journals/pacmpl/MuylderND24}, an implementation
of the CH binary PTT. The nullary and binary variants operate in a similar way.
The pseudo-code we write uses syntax similar to Agda.
We omit universe levels,
and $\multimap$ is parsed like $\to$, e.g., it is right associative.
We use $;$ to group several declarations in one line.

Lastly we indicate that the SAP holds at equality types
$(a'_0\,a'_1 : \aint \multimap A) \to ((x:\aint)\multimap a_0'x \equiv_A a_1'x) \simeq a'_0 \equiv a'_1$.
Proofs of this fact in the binary case can be found in~\cite{DBLP:journals/lmcs/CavalloH21, DBLP:journals/pacmpl/MuylderND24}.
The nullary proof is obtained by erasing all mentions of (bridge) endpoints.

\subparagraph*{The extent primitive}
The rule for the extent primitive ($\textsc{ext}$)
together with the rules of $\multimap$ and $\cU$ provide
a term $\sfext$ in the empty context with the following type.
\begin{lstlisting}[mathescape]
$\sfext$ : {A : $\aint$ $\multimap$ $\cU$} {B : (x:$\aint$) $\multimap$ A x $\to$ $\cU$}
  (f' : (a' : (x:$\aint$) $\multimap$ A x) $\to$ (x:$\aint$) $\multimap$ B x (a' x)) $\to$
  (x:$\aint$) $\multimap$ (a : A x) $\to$ B x a
\end{lstlisting}

From a function $f'$ mapping a bridge in $A$ to a bridge in $B$,
the extent primitive lets us build a bridge in the dependent function type formed from $A$ and $B$, or $\Pi\,A\,B$ for short.
Concisely put, the extent primitive validates one direction of the SAP at $\Pi$.
It can be upgraded into the following equivalence, using the $\beta$-rule of extent, described below
(as a side note, the proof also uses a lemma akin to a propositional $\eta$-rule for extent).
\begin{equation*}
((a' : (x:\aint) \multimap A\,x) \to (x:\aint)\multimap B\,x\,(a'\,x))
\quad \simeq \quad
(x : \aint) \multimap (a : A\,x) \to B\,x\,a
\end{equation*}
By way of comparison, this equivalence appears in~\cite{DBLP:journals/entcs/PittsMD15} (Example 2.2)
though the left-to-right direction is implemented via
a composite capturing operation instead of a primitive like extent.
Proofs of the above equivalence
in the binary case can be found in~\cite{DBLP:journals/lmcs/CavalloH21, DBLP:journals/pacmpl/MuylderND24}.
The nullary proof is obtained by erasing all mentions of endpoints.
This equivalence entails $((\aint \multimap A) \to (\aint \multimap B)) \simeq \aint \multimap (A \to B)$ in
the non-dependent case.


The $\beta$-rule of extent (\textsc{ext$\beta$}) is peculiar
because a substituted premise $a[x/y]$ appears in the redex
on the left-hand side.
To compute the right-hand side out of the left-hand side only,
the premise $a$ is rebuilt (*) and a term where a variable is bound in $a$
is returned.
The step (*) is possible thanks to the restriction in the context $\Gamma\restr x, (y:\aint)$ of $a$.
This is quite technical and explained in~\cite{DBLP:journals/lmcs/CavalloH21, DBLP:journals/pacmpl/MuylderND24}.
We instead explain what that entails for our $\sfext$ program above.
Non-trivial examples
of $\beta$-reductions for extent will appear in \cref{sec:name-ind}.

In a context $\Gamma$ containing $(x:\aint)$, the term $\Gamma\vdash \sfext\,f'\,x\,a$ reduces if $\Gamma \vdash a$ is a term that does not mention cartesian variables
strictly to the right of $x$ in $\Gamma$, i.e.\ declared after $x$ in $\Gamma$. In particular $a$ can mention $x$.
If such a freshness condition holds, $x$ can in fact be soundly captured in $a$ and the reduction can trigger.
By default the reduction does not trigger because $f', x, a \vdash \sfext \,f'\,x\,a$ does not
satisfy the freshness condition as in this case $a$ is a
variable appearing to the right of $x$ in the context.

\subparagraph*{Gel types}
Internally parametric type theories that feature interval-based
Bridge types~\cite{moulin2016internalizing, DBLP:journals/lmcs/CavalloH21}
need a primitive to convert an $n$-ary relation of types into an $n$-ary bridge.
In the CH theory the primitive is called $\sfGel$.
The rules for $\sfGel$ types together with the rules of $\multimap$ and $\cU$ imply
the existence of the following programs in the empty context. Note
that $\sfung$ is called ungel in the CH theory.
\begin{lstlisting}[mathescape]
$\sfGel$ : $\cU$ $\to$ $\aint$ $\multimap$ $\cU$
$\sfgel$ : {A : $\cU$} $\to$ A $\to$ (x : $\aint$) $\multimap$ $\sfGel$ A x
$\sfung$ : {A : $\cU$} $\to$ ((x : $\aint$) $\multimap$ $\sfGel$ A x) $\to$ A
\end{lstlisting}

Similar to extent, $\sfGel$ validates one direction of the SAP, this time at the universe $\cU$.
The $\sfGel$ function can be upgraded into an equivalence $\cU  \simeq  (\aint \multimap \cU)$
where $\sfGel^{-1}\,A' := (x:\aint)\multimap A'\,x$,
and by using the rules of $\sfGel$ and $\sfext$, and notably univalence.
The proof also requires showing that $\mathsf{gel}$ and $\sfung$
are inverses, i.e.\ the SAP at $\sfGel$ types $A \simeq (x:\aint) \multimap \sfGel\,A\,x$.
Again these theorems are proved in~\cite{DBLP:journals/lmcs/CavalloH21, DBLP:journals/pacmpl/MuylderND24} in the binary
case, and the nullary proofs are obtained by erasing endpoints.

From a nominal perspective, we observe that $\sfGel$
is a type former that can be used to express freshness information.
This can be seen by looking at the \textsc{GelI} rule: canonical inhabitants
of $\sfGel\,A\,x$ are equivalently terms $a:A$ such that $x$ is fresh in $a$.
Conversely the $\sfung$ primitive is a binder that makes available a fresh variable $x$
when a value of type $A$ is being defined, as long as the result value is typally fresh w.r.t. $x$.
The $\sfung$ primitive can also be used to
define a map forgetting freshness information.
\begin{lstlisting}[mathescape]
$\mathsf{forg}$ : {A : $\cU$} $\to$ (x:$\aint$) $\multimap$ $\sfGel$ A x $\to$ A
$\mathsf{forg}$ {A} = $\mathsf{ext}$ [$\lambda$(g':(x:$\aint$) $\multimap$ $\sfGel$ A x).$\lambda$(_:$\aint$). $\mathsf{ung}$ g']
\end{lstlisting}
To our knowledge the equivalence $\cU  \simeq  (\aint \multimap \cU)$
had not been considered in the literature on nominal techniques.
Lastly, \textsc{Gel$\eta$} can be used if a certain freshness side condition holds,
similar to \textsc{ext$\beta$}. We will not need to make explicit use of \textsc{Gel$\eta$}.

\subsection{The $\nm$ type, Name Induction, Nominal Data Types}\label{sec:name-ind}

So far the rules we presented involved occurrences of the bridge interval
morally to the left of $\vdash$,
as affine comprehensions.
Since we wish to use nullary PTT as a nominal framework, we need a first-class
type of names $\vdash \nm \: \sftype$. Indeed this is needed to express
nominal data types, whose constructors may take names as arguments.
For example the $\mathsf{var}$ constructor of the $\Ltm$ type declared
in \cref{sec:intro} has type $\mathsf{var} : \nm \to \mathsf{Ltm}$, a regular ``cartesian'' function type.
Constructors must be cartesian functions because
that is what the initial algebra semantics of (even nominal) data types dictates.
Besides, it is unclear whether there exists
a sound notion of data types $D$ with ``constructors'' of the form $\aint \multimap D$.


Perhaps the most important result in our system is the SAP at the type of names $\nm$
which reads $1 + \nm \simeq (\aint \multimap \nm)$.
The principle expresses that a bridge at the type of names $\nm$ is either the ``identity'' bridge $\cart : \aint \multimap \nm$ or a constant bridge.
This principle is proved based on our rules for $\nm$, explained now.
The rules of $\nm$ together with the other rules of \cref{fig:nullary-ptt-rules}
entail the existence of the following programs in the empty context.
\begin{lstlisting}[mathescape]
$\nm$ : $\cU$
$\cart$ : $\aint$ $\multimap$ $\nm$
$\indN$ : {B : $\nm$ $\to$ $\cU$} $\to$
  (x:$\aint$) $\multimap$ (n:$\nm$) $\to$
  (b0 : B ($\cart$ x)) $\to$ (b1 : (g : $\sfGel$ $\nm$ x) $\to$ B (forg x g)) $\to$ B n
\end{lstlisting}

The \textsc{$\nm$I} rule expresses that the canonical inhabitants of $\nm$
in a context $\Gamma$ are simply the affine variables $(x:\aint)$ appearing in $\Gamma$.
The ``identity'' bridge program $\cart$ above is derived from that rule.
In simple terms, $\cart$ coerces affine bridge variables $x$ into names $\cart\,x : \nm$.

\subparagraph*{Name induction}
The $\indN$ program/rule is an induction principle, or dependent eliminator
for the type $\nm$. 
It expresses that in a context $\Gamma$ containing
an affine variable $(x:\aint)$ we can do a case analysis on a term $\Gamma \vdash n:\nm$.
Either $x$ is bound in $n$ and $n$ is in fact exactly $\cart\,x$,
or $x$ is \emph{fresh} in $n$. The call $\indN\,x\,n\,b_0\,b_1$ returns
$b_0$ if $n=\cart\,x$ (see rule \textsc{$\nm\beta_0$})
and returns $b_1\,(\sfgel\,n\,x)$ if $x$ is fresh in $n$ (see rule \textsc{$\nm\beta_1$}).
The freshness assumption in $b_1$ is expressed typally using a $\sfGel$ type.

We show that the rules $\nm\beta_0$ and $\nm\beta_1$ are well-typed, i.e.\ that
the $\indN$ program above reduces to something of type $B\,n$ in both scenarios.
If $n = \cart\,x$ then the reduction result is $\indN \,x\,(\cart\,x)\,b_0\,b_1 = b_0 : B\,(\cart\,x)$
and $B(\cart\,x) = B\,n$.
Else if $x$ is fresh in $n$ then $\sfgel\,n\,x$ typechecks and
the reduction result is $\indN \,x\,n\,b_0\,b_1 = b_1\,(\sfgel\,n\,x) : B (\mathsf{forg} \,x \,(\sfgel\,n\,x))$
where $\mathsf{forg}$ is the function defined in \cref{sec:nullary-ch}.
And we have:
\begin{align*}
\mathsf{forg}\,x\,(\sfgel\,n\,x) &= \sfext \, [\lambda (g':(x:\aint) \multimap \sfGel\,\nm\,x). \lambda (\_ : \aint).\sfung \,g'] \, x \,(\sfgel\,n\,x) &(\text{def.})\\
        &= (\lambda (\_ : \aint).\,\sfung\,(\lambda y.\, \sfgel \,n\,y)) \, x                                                             &(\textsc{ext}\beta)\\
        &= \sfung\,(\lambda y.\, \sfgel \,n\,y)                                                                                          = n                                                                                                                              &{}\hspace{-2em}({\multimap}\beta, \textsc{Gel}\beta)
\end{align*}
The \textsc{ext$\beta$} rule triggers because (1) by assumption, $x$ is fresh in $n$,
i.e.\ $n$ does not mention $x$, nor cartesian variables strictly to the right of $x$
(2) thus the term $\sfgel\,n\,x$ mentions $x$ but no cartesian variables declared later.
The latter condition is exactly the condition under which this $\sfext$ call can reduce.
To that end, $x$ is captured in the term $\sfgel\,n\,x$ leading to a term $g' := \lambda y.\,\sfgel\,n\,y$
appearing on the second line.

\subparagraph*{SAP at $\nm$}
We now prove that $1+\nm \simeq \aint \multimap \nm$.
We take inspiration from~\cite{DBLP:journals/lmcs/CavalloH21} who developed a
relational encode-decode technique to prove the
SRP at data types. The type $\nm$ is defined via an induction principle,
and a similar technique can be applied.
\begin{lstlisting}[mathescape]
$\mathsf{decode}$ : 1 + $\nm$ $\to$ $\aint$ $\multimap$ $\nm$
$\mathsf{decode}$ (inl tt) = $\lambda$x. $\cart$ x   ;   $\mathsf{decode}$ (inr n) = $\lambda$_. n

t1 : ($\aint$ $\multimap$ $\nm$) $\to$ (x:$\aint$) $\multimap$ 1 + $\sfGel$ $\nm$ x
t1 n' x = $\indN$ x (n' x) (inl tt) ($\lambda$ (g:$\sfGel$ $\nm$ x). inr g)

t2pre : (x:$\aint$) $\multimap$ (1 + $\sfGel$ $\nm$ x) $\to$ $\sfGel$ (1 + $\nm$) x
t2pre x (inl tt) = $\sfgel$ (inl tt) x
t2pre x (inr g) = $\sfext$ [$\lambda$g'. $\lambda$y. $\sfgel$ (inr ($\sfung$ g')) y] x g

t2 : ((x:$\aint$) $\multimap$ 1 + $\sfGel$ $\nm$ x) $\to$ (x:$\aint$) $\multimap$ $\sfGel$ (1 + $\nm$) x
t2 = $\lambda$ s' x. t2pre x (s' x)

$\mathsf{encode}$ : ($\aint$ $\multimap$ $\nm$) $\to$ 1 + $\nm$
$\mathsf{encode}$ n' = $\sfung$ (t2 (t1 n'))
\end{lstlisting}

It remains to prove the roundtrip equalities.
The roundtrip for $s: 1+\nm$ is obtained by induction on $s$.
If $s = \mathsf{inl}{\,\mathsf{tt}}$ then $(\sfung\circ t_2\circ t_1\circ\mathsf{decode})\,s =
(\sfung\circ t_2\circ t_1) (\lambda x.\,\cart\,x) \overset{\nm\beta_0}{=}
(\sfung\circ t_2) (\lambda x.\,\mathsf{inl}\,\mathsf{tt}) = \sfung\,(\lambda x.\,\sfgel\,(\mathsf{inl}\,\mathsf{tt})\,x) = \mathsf{inl}\,\mathsf{tt} = s$.
If $s = \mathsf{inr}\,n$ then
$(\sfung\circ t_2\circ t_1\circ\mathsf{decode})\,s = (\sfung\circ t_2\circ t_1)(\lambda x.\,n)
\overset{\nm\beta_1}{=} (\sfung\circ t_2)(\lambda x.\,\mathsf{inr}(\sfgel\,n\,x))
= \sfung\, (\lambda x.\, \sfext\,[\lambda g' y.\,\sfgel\,(\mathsf{inr}(\mathsf{ung}\, g'))\,y]\,\allowbreak  x \,(\sfgel\,n \,x))
\overset{\textsc{ext}\beta}{=} \sfung\,(\lambda x.\, \sfgel\,(\mathsf{inr}\,n) x) = \mathsf{inr}\,n = s$.

The other roundtrip is the last type in the following chain of equivalences.
The first type is a sufficient condition that can be understood as a propositional $\eta$-rule for $\nm$.
\begin{align*}
(x:\aint) \multimap (n:\nm) \to (n \equiv_\nm \sfext\,[\mathsf{decode}\circ \mathsf{encode}]\,x \, n) &\qquad\simeq_{(\text{SAP}_\to)} \\
(n' : \aint \multimap \nm) \to (x:\aint) \multimap (n' x \equiv_\nm (\mathsf{decode}\circ \mathsf{encode})\,n'\,x) &\qquad\simeq_{(\text{SAP}_\equiv)}\\
(n' : \aint \multimap \nm) \to n' \equiv (\mathsf{decode}\circ \mathsf{encode}) \,n'
\end{align*}
The $\sfext$ in the first line vanishes in the second because \textsc{ext$\beta$} triggers.
We prove the sufficient condition.
Let $(x:\aint), (n:\nm)$ in context.
We reason by name induction.
If $n = \cart\,x$ then the right-hand side
is $\sfext\,[\mathsf{decode}\circ \mathsf{encode}]\,x \, (\cart\,x)
\overset{\textsc{ext}\beta}{=} ((\mathsf{decode}\circ \mathsf{encode})\,\cart)\,x$.
From the proof above we know that $\mathsf{encode}\,\cart = \mathsf{inl}\,\mathsf{tt}$
, and by definition $\mathsf{decode}\,(\mathsf{inl}\,\mathsf{tt}) = \cart$.
So both sides of the equality are equal to $\cart\,x$.
Else, formally we must provide a $b_1$ argument to $\mathsf{ind}_\nm$.
We define $b_1 := b'_1\,x$ where the type of $b'_1$ is the first type in this SAP derivation
(performed in the empty context and where $[\ldots] := \mathsf{decode}\circ \mathsf{encode}$):
\begin{align*}
 &\quad\,\,  (x:\aint) \multimap (g : \sfGel\,\nm\,x) \to \mathsf{forg}\,x\,g \equiv \sfext\,[\ldots]\,x\,(\mathsf{forg}\,x\,g)\\
 &\simeq     (g' : (x:\aint) \multimap \sfGel\,\nm\,x) \to (x:\aint) \multimap \mathsf{forg}\,x\,(g'\,x) \equiv \sfext\,[\ldots]\,x\,(\mathsf{forg}\,x\,(g'\,x))\\
 &\simeq     (n : \nm) \to (x:\aint) \multimap \mathsf{forg}\,x\,(\sfgel\,n\,x) \equiv \sfext\,[\ldots]\,x\,(\mathsf{forg}\,x\,(\sfgel\,n\,x))
\end{align*}
By a computation above,
$\mathsf{forg}\,x\,(\sfgel\,n\,x) = n$ since $x$ is fresh in $n$.
The right-hand side computes to $n$ as well by $\textsc{ext}\beta$ and definition of $[\ldots] = \mathsf{decode}\circ \mathsf{encode}$.

\subparagraph*{Nominal data types}
When looking at a specific example of a nominal data type $D$
we temporarily extend the type system with the rules of $D$.
This includes the dependent eliminator of $D$. For example
in the case of the $\Ltm$ type of \cref{sec:intro} we add a rule
that entails the existence of this closed program:
\begin{lstlisting}[mathescape]
indLtm : (P : $\Ltm$ $\to$ $\cU$) $\to$ ((n : $\nm$) $\to$ P(var n)) $\to$
($\forall$a b. P a $\to$ P b $\to$ P(app a b)) $\to$
($\forall$g. ((x:$\aint$) $\multimap$ P(g x)) $\to$ P(lam g)) $\to$ $\forall$t.P t
\end{lstlisting}

\subsection{The Structure Abstraction Principle (SAP)}
\label{sec:struct-abstr-princ}

\begin{table}
\noindent\begin{tabular}{ | p{2.3cm} | p{4.5cm} || c || c c |}
\hline
Parameters & $K'$ & $\simeq$ & \multicolumn{2}{c|}{$(x : \aint) \multimap K$}\\
\hline\hline
    \multirow{2}{2.3cm}{$A : \aint \multimap \cU$, $B~:~(x:~\aint)~\multimap$  $A~x~\to\cU\ldotp $}
& $(a' : (x:\aint) \multimap A\,x) \to (x:\aint)\multimap B\,x\,(a'\,x)$ & $\simeq$ & $(x : \aint) \multimap$ & $(a : A~x) \to B~x~a$\\
\cline{2-5}
& $(a':(x : \aint) \multimap A~x)\times((x : \aint) \multimap B~x~(a'~x))$ & $\simeq$ & $(x : \aint) \multimap$ & $(a : A~x)\times (B~x~a)$\\
\hline
& $\cU$ & $\simeq$ & $(x : \aint) \multimap$ & $\cU$\\
  \hline
$A:\cU, a_0, a_1 : \aint \multimap A\ldotp $& $ a_0 \: \equiv_{\aint \multimap A} \:a_1$ & $\simeq$ & $(x : \aint) \multimap$ & $a_0~x \equiv a_1~x$\\
\hline
& $1+\nm$ & $\simeq$ & $(x : \aint) \multimap$ & $\nm$\\
  \hline
$A : \cU$& $A$ & $\simeq$ & $(x : \aint) \multimap$ & $\sfGel\,A\,x$\\
  \hline
$A : \cU, y \neq x$& $\sfGel\,((x : \aint) \multimap A)\,y$ & $\simeq$ & $(x : \aint) \multimap$ & $\sfGel\,A\,y$\\
  \hline
$A : (x:\aint)\multimap (y:\aint)\multimap \cU$& $(y:\aint)\multimap (x:\aint)\multimap A\,x\,y$ & $\simeq$ & $(x : \aint) \multimap$ & $(y:\aint)\multimap A\,x\,y$\\
  \hline
\end{tabular}
\caption{The Structure Abstraction Principle.}
\label{tbl:sap-table}
\end{table}


\subparagraph*{For types and terms}
As hinted above,
PNTT validates the Structure Abstraction Principle (SAP).
Similar to the SIP of HoTT/UF, or the SRP of binary PTT~\cite{DBLP:journals/pacmpl/MuylderND24},
the SAP defines how each type former commutes with the (nullary) Bridge type former.
\cref{tbl:sap-table} lists several SAP instances, including the ones we have encountered so far.
Each instance is\footnote{Orienting
$\mathsf{SAP}_K$ this way enables $\mathsf{SAP}_\cU = \sfGel$ and $\mathsf{SAP}_\Pi = \sfext$.}
of the form $\forall \cdots\ldotp K' \simeq ((x : \aint)\multimap K)$,
where $K,K'$ may depend on some parameters.
Note that the instances in \cref{tbl:sap-table} involve primitive type formers $K$.
In order to obtain the SAP instance of a composite type $K$,
we can combine the SAP instances of the primitives used to define $K$
(this was done \EG{} in \cref{sec:name-ind} when proving the propositional $\eta$-rule of $\nm$,
and is further discussed below).
Types of DTT can contain terms and accordingly there exists a SAP for terms.
Let $\mathsf{SAP}_K : K' \simeq (\aint \multimap K)$ be a SAP instance.
The SAP of the term $k:K$ is a pair $(k', p)$ where $k' : K'$
and $p : \mathsf{SAP}_K\,k' \equiv \lambda(\_:\aint).k$, \IE{}
$k'$ is the reflexivity bridge $\mathsf{rfl}\,k := \lambda(\_:\aint).k$ of $k:K$ through the $\mathsf{SAP}_K$ equivalence,
up to propositional equality.

\subparagraph*{Relationship to the translation}
For types, it is basically the case that the type $K'$ provided by the SAP at $K$ is
$[K]_0 : \cU$, the recursively and externally defined nullary parametricity translation of $K$. The $n$-ary translation is briefly discussed in \cref{sec:intro}.
Note that in practice it is sometimes useful to consider SAP instances where $K'$ is a type different but equivalent to $[K]_0$, but this
detail can be ignored for clarity.
For terms, it is also the case that the SAP at $k:K$ provides a $k'$ which is $[k]_0 : [K]_0$, the recursively defined nullary translation of $k:K$
(analogously it is sometimes useful to let $k'$ be a term different but propositionally equal to $[k]_0$).
From that perspective, $\mathsf{SAP}_K : [K]_0 \simeq \aint \multimap K$ says that the Bridge type of $K$
is its translation up to an equivalence, and the SAP of a term $k:K$ says that $[k]_0 \equiv \mathsf{SAP}_K^{-1}\,(\lambda(\_:\aint).k)$, \IE{}
the reflexivity bridge at $k:K$ is its translation up to a propositional equality.
It is useful to give a name to the latter right-hand side.
If a SAP instance is fixed for $K$ we define the \emph{observational parametricity} of a term $k:K$ to be $[k]_0^{\mathcal{O}} := \mathsf{SAP}_K^{-1}\,(\lambda(\_:\aint).k)$.
The various parametricity mappings for types and terms can be informally summarized as follows, where $\overset{m}{\to}$ denotes a metatheoretical function.
\begin{align*}
[-]_0 : \cU \overset{m}{\to} \cU &\quad;\quad [-]_0, [-]_0^{\mathcal{O}}  : \lbrace K : \cU \rbrace \overset{m}{\to} K \overset{m}{\to} [K]_0\\
\aint \multimap {-} : \cU \to \cU &\quad;\quad \mathsf{rfl} : \lbrace K : \cU \rbrace \to (k:K) \to \aint \multimap K\\
\mathsf{SAP} : (K:\cU) \overset{m}{\to} [K]_0 \simeq \aint \multimap K
     &\quad;\quad \mathsf{SAP} : \lbrace K:\cU \rbrace \overset{m}{\to} (k:K) \overset{m}{\to} [k]_0 \equiv [k]_0^{\mathcal{O}}
\end{align*}

As an example, we compute the observational parametricity of a function $f : A \to B$.
Suppose $A$ and $B$ are types with SAP instances $A' \simeq (\aint \multimap A)$ and
$B' \simeq (\aint \multimap B)$.
The SAP at $A \to B$ is the composition
$\mathsf{SAP}_{A\to B}^{-1} :(\aint \multimap (A \to B)) \: \simeq\: ((\aint \multimap A) \to \aint \multimap B) \: \simeq \: A' \to B'$.
The map $\mathsf{SAP}_{A\to B}^{-1}$ is given by $\lambda f'.\,\mathsf{SAP}_B^{-1}\circ (\mathsf{ext}^{-1}\,f') \circ \mathsf{SAP}_A$
where $\sfext^{-1} f'= \lambda(a' : \aint \multimap A).\,\lambda(x:\aint).\,f'\,x\,(a'\,x)$.
With this choice of SAP instances,
we have $[f]_0^{\mathcal{O}} : A' \to B'$ and $[f]_0^{\mathcal{O}} = \mathsf{SAP}_B^{-1}\circ (\aint \multimap f) \circ \mathsf{SAP}_A$ where
$(\aint \multimap f) = \lambda a'\,x.\,f\,(a' x)$ is the action of $f$ on bridges.
So the SAP for a function $f : A\to B$ asserts that its action on bridges agrees with its translation, up to the SAP at $A,B$.


The SAP of a composite type $K$ is obtained
by \emph{manually} applying the SAP rules from Table~\ref{tbl:sap-table}.
This process will produce a translation $K'$
that is equivalent to $\aint \multimap K$.
In the binary case, we have shown in earlier work~\cite{DBLP:journals/pacmpl/MuylderND24} that
this process can be systematized, by organizing types and terms together with their SRP instances into a (shallowly embedded) type theory
called ROTT.
For types $K$ and terms $k : K$ that fall in the ROTT syntax, the parametricity
translations and SAP instances can be derived straightforwardly.
Although we believe the same process applies here, we have not constructed the corresponding DSL,
instead applying SAP instances manually in examples.

\subsection{About Semantics}\label{sec:semantics-soundness}

Similar to our theory PNTT, our model is a simple extension of
Cavallo and Harper's presheaf model~\cite{DBLP:journals/lmcs/CavalloH21} with the arity of parametricity changed from $2$ to $0$.
This model is $\mathsf{Psh}(\boxslash_2 \times \Box_0)$, where
$\boxslash_2$ is the binary cartesian cube category~\cite{DBLP:journals/mscs/AngiuliBCHHL21} for path dimensions,
and $\Box_0$ is the $0$-ary affine cube category for bridge dimensions.
Concretely $\Box_0$ is the opposite of the category of finite ordinals and injections.
We keep this section overall succinct, not for lack of formal understanding, because these semantics are a well-understood combination of
established models of type theory~\cite{Hofmann97,psh-universes,DBLP:journals/mscs/AngiuliBCHHL21,bernardy2015presheaf,DBLP:journals/lmcs/CavalloH21}.


In this model the type $\nm$ is interpreted as the Yoneda embedding of the base object with 1 bridge dimension and no path dimensions.
The elimination and computation rules for this type are based on a semantic isomorphism $x : \aint \vdash \nm \cong 1 + \sfGel\,\nm\,x$ which is straightforwardly checked.
In fact, the above (taken as an equivalence) is equivalent (by several invocations of the SAP) to $(\aint \multimap \nm) \simeq 1+\nm$, an equivalence semantically
proven by Hofmann~\cite{DBLP:conf/lics/Hofmann99}.
Kan fibrancy of $\nm$ is semantically trivial, since we can tell from the base category that any path in $\nm$ will be constant.
As for computation of Kan operations at $\nm$, we propose to wait until all arguments reduce to $\cart\,x$ for the same affine name $x$, in which case we return $\cart\,x$.
This is in line with how Kan operations for positive types with multiple constructors are usually reduced~\cite{cohen_cubical_2016,DBLP:journals/mscs/AngiuliBCHHL21}.
Regarding nominal data types, which we only consider through examples in this paper, we take the viewpoint that these arise from an interplay between the usual type formers, bridge types, $\nm$, and an initial algebra operation for `nominal strictly positive functors'.
We do not attempt to give a categorical description of such functors or prove that they have initial algebras.
The Kan operation will reduce recursively and according to the Kan operations of all other type formers involved.

We note that $\Box_0$ is the base category (carrier of the site) of the Schanuel sheaf topos~\cite[\S 6.3]{pitts-nominal-sets} which is equivalent to the category of nominal sets~\cite{pitts-nominal-sets}, used to model FreshMLTT~\cite{DBLP:journals/entcs/PittsMD15}.
The sheaf condition requires that presheaves $\Gamma : \Box_0\op \to \rmSet$ preserve pullbacks, i.e.\ compatible triples in $\Gamma_{U \uplus V} \to \Gamma_{U \uplus V \uplus W} \leftarrow \Gamma_{U \uplus W}$ have a unique preimage in $\Gamma_U$.
Conceptually, if a cell $\gamma \in \Gamma_{U \uplus V \uplus W}$ is both fresh for $V$ and for $W$, then it is fresh for $V \uplus W$, with unique evidence.
This property is not available internally in nullary PTT, nor do we have the impression that it is important to add it, so we content ourselves by modeling types as \emph{pre}sheaves over $\Box_0$.

\section{Nominal Primitives for Free}
\label{sec:nomin-prim-free}

In this section, we argue that the central features in a number of earlier nominal dependent type systems can essentially be recovered in nullary PTT.
One notable exception is nominal pattern matching (discussed in the next section), which is recovered in our extension PNTT.
Concretely, we consider here Shinwell, Pitts and Gabbay's FreshML~\cite{DBLP:conf/icfp/ShinwellPG03}, Sch\"opp and Stark's bunched
nominal type theory \sysSchopp{}~\cite{schopp2004dependent,DBLP:phd/ethos/Schopp06}, Cheney's \sysCheney{}~\cite{DBLP:journals/corr/abs-1201-5240} and Pitts, Matthiesen and Derikx's FreshMLTT~\cite{DBLP:journals/entcs/PittsMD15}.
The central features we identify there, are: existential and universal name quantification, a type former expressing freshness for a given name, name swapping, and locally scoped names.
Existential and universal name quantification are known to be equivalent in the usual (pre)sheaf or nominal set semantics of nominal type theory, but generally have quite different typing rules: the former is an existential type former with pair-like constructor and matching eliminator (opening the door to matching more deeply), whereas the latter is a universal type operated through name abstraction and application (getting in the way of matching more deeply).
We note that some systems (FreshMLTT, \sysCheney{}) support multiple name types, something we could also easily accommodate, but leave out so as not to distract from the main contributions. \sysSchopp{} even allows substructural quantification and typal freshness for arbitrary closed types (rather than just names), which is not something we intend to support and which inherently seems to require a bunched context structure.
In what follows, we will speak of `our rules', not to claim ownership (as they are inherited from Bernardy, Coquand and Moulin~\cite{bernardy2015presheaf} and Cavallo and Harper~\cite{DBLP:journals/lmcs/CavalloH21}), but to distinguish them from the other systems.

\subparagraph{Universal name quantification}
Universal name quantification is available in \sysSchopp{}, \sysCheney{} and FreshMLTT. In our system, it is done using the nullary bridge type $(x : \aint) \multimap A$, whose rules are given in \cref{fig:nullary-ptt-rules}.
The rules $\multimap$F and $\multimap$I correspond almost perfectly with the other three systems; for \sysSchopp{} we need to keep in mind that our context extension with a fresh name, is semantically a monoidal product.
The application rules in \sysCheney{} and \sysSchopp{} also correspond almost precisely to $\multimap$E, but the one in \sysSchopp{} is less algorithmic than ours.
Specifically, the BNTT bunched application rule takes a function $\Theta \vdash f : (x : T) \multimap A\,x$ and an argument $\Delta \vdash t : T$ and produces $f\,t : A\,t$ in a non-general context $\Theta * \Delta$. Our rule
$\multimap$\textsc{E} (inherited from~\cite{bernardy2015presheaf,DBLP:journals/lmcs/CavalloH21}) improves upon this by taking in an arbitrary context $\Gamma$ and \emph{computing} a context $\Gamma\restr x$ such that there is a morphism $\Gamma \to (\Gamma\restr x, (x : \aint))$.
The application rule in FreshMLTT is similar, but uses definitional freshness (based on a variable swapping test) to ensure that the argument is fresh for the function.

\subparagraph{Typal freshness} A type former expressing freshness is available in \sysSchopp{} (called the free-from type).
FreshMLTT uses definitional freshness instead. In nullary PTT, elements of $A$ for which $x$ is fresh are classified by the type $\sfGel \, A \, x$.
\sysSchopp{}'s free-from types only apply to closed types $A$, so \textsc{GelF} is more general.
\sysSchopp{}'s introduction rule corresponds to \textsc{GelI}, which is however again more algorithmic.
\sysSchopp{}'s elimination rule is explained in terms of single-hole bunched contexts~\cite[\S 4.1.1]{DBLP:phd/ethos/Schopp06} which specialize in our setting (where the only monoidal product is context extension with a name) to contexts with a hole up front. Essentially then, the BNTT rule can be phrased in nullary PTT as
\begin{center}
	\AxiomC{$\Gamma, x : \aint, z : \sfGel\,B\,x, \Theta \vdash T \, \sftype$}
	\AxiomC{$\Gamma, y : B, x : \aint, \Theta[\sfgel\,y\,x/z] \vdash t : T[\sfgel\,y\,x/z]$}
	\BinaryInfC{$\Gamma, x : \aint, z : \sfGel\,B\,x, \Theta \vdash t' : T$}
	\DisplayProof
\end{center}
where $\Gamma$ must be empty, together with $\beta$- and $\eta$-rules establishing that the above operation is inverse to applying the substitution $[\sfgel\,y\,x/z]$.\footnote{The original rule immediately subsumes a substitution $\Delta \to (x : \aint, z : \sfGel\,A\,x)$.}
We can in fact accommodate the rule for non-empty $\Gamma$.
Without loss of generality, we can assume $\Theta$ is empty: by abstraction/application, we can subsume $\Theta$ in $T$.\footnote{This may
raise questions about preservation of substitution w.r.t.\ $\Theta$. However, it is not at all standard for dependently typed elimination rules to guarantee preservation of substitution w.r.t.\ a telescope depending on the eliminee.}
We then have an equivalence
{\small
\begin{align*}
	(y : B) \to ((x : \aint) \multimap T[\sfgel\,y\,x/z])
	&\quad \simeq \quad
	(y : (x : \aint) \multimap \sfGel\,B\,x) \to ((x : \aint) \multimap T[y\,x/z]) \\
	&\quad \simeq \quad
	(x : \aint) \multimap (z : \sfGel\,B\,x) \to T
\end{align*}%
}%
where in the first step, we precompose with the $\sfgel$/$\sfung$ isomorphism (the SAP for $\sfGel$), and in the second step, we apply the $\sfext$ equivalence (the SAP for functions).

\subparagraph{Existential name quantification}
Existential name quantification is available in FreshML and \sysSchopp{}.
We first discuss how we can accommodate the \sysSchopp{} rules, and then get back to FreshML.
The \sysSchopp{} existential quantifier is just translated to the nullary bridge type $(x : \aint) \multimap A$ again, i.e.\ both quantifiers become definitionally the same type in nullary PTT.
\sysSchopp{}'s introduction rule follows by applying a function
$\sfbind : (x : \aint) \multimap B\,x \to \sfGel\,((w : \aint) \multimap B\,w)\,x$
which is obtained from the identity function on $(w : \aint) \multimap B\,w$ by the SAP for functions and $\sfGel$.
\sysSchopp{}'s elimination rule essentially provides a function
$\mathsf{matchbind} : \paren{ (x : \aint) \multimap B\,x \to \sfGel\,C\,x } \to \paren{ ((x : \aint) \multimap B\,x) \to C }$
such that if we apply $\mathsf{matchbind}\,f$ under $\sfGel$ to $\sfbind\,x\,b$, then we obtain $f\,x\,b$.
Again, the SAP for $\sfGel$ and functions reveals that the source and target of $\mathsf{matchbind}$ are equivalent.

The non-dependently typed system FreshML has a similar type former, but lacks any typal or definitional notion of freshness.
Their introduction rule then follows by postcomposing the above $\sfbind$ with the function $\mathsf{forg}$ that forgets freshness (\cref{sec:nullary-ch}).
If we translate FreshML's declarations $\Gamma \vdash d : \Delta$ to operations that convert terms $\Gamma, \Delta \vdash t : T$ to terms $\Gamma \vdash t\accol{d}$ (not necessarily by substitution), then we can translate their declaration $\Gamma \vdash \mathtt{val}\,\angles{x} y = e : (x : \aint, y : B)$, where $e$ is an existential pair, as $\mathsf{matchdecl}\,e$ where
$\mathsf{matchdecl} : \paren{ e : \aint \multimap B } \to \paren{ t : \aint \multimap B \to T } \to \paren{ \aint \multimap T }$.
This function is obtained by observing that the second argument type, by the SAP for functions, is equivalent to $(\aint \multimap B) \to (\aint \multimap T)$.
It may be surprising that the result has type $\aint \multimap T$ rather than just $T$; this reflects the fact that FreshML cannot enforce freshness, and is justified by the fact that it allows arbitrary declaration of fresh names via the declaration $\Gamma \vdash \mathtt{fresh}\,x : (x : \aint)$, which we do not support.

\subparagraph{Swapping names}
The name swapping operation is available in FreshML and FreshMLTT.
We can accommodate the full rule of FreshML (which is not dependently typed) and a restricted version of the rule in FreshMLTT, where we allow the type of the affected term to depend on the names being swapped, but other than that, only on variables fresh for those names.
We simply use the function $\sfswap : (x\,y : \aint) \multimap T\,x\,y \to T\,y\,x$ whose type is equivalent to $\paren{(x\,y : \aint) \multimap T\,x\,y} \to \paren{(x\,y : \aint) \multimap T\,y\,x}$ by the SAP, and the latter is clearly inhabited by $\lambda\,t\,x\,y.\,t\,y\,x$.
Note that name swapping needs to happen w.r.t.\ affine names: swapping the cartesian names $x, y : \nm$ in the term $\sfapp\,(\sfapp\,(\sfvar\,x)\,(\sfvar\,y))\,(\sfvar\,z) : \sfLtm$ does not commute with contraction $-[x/x, x/z]$.
Finally, we note that the aforementioned restriction on the type can be mitigated in an ad hoc manner, because types $T : (x\,y : \aint) \multimap \Delta \to \cU$ (where $\Delta$ denotes any telescope consisting of both affine names and non-affine variables) are by the SAP in correspondence with types $\Delta'' \to (x\,y : \aint) \multimap \cU$ for a different telescope $\Delta''$.
This however does not imply that we can accommodate the full FreshMLTT name swapping rule in a manner that commutes with substitution.
We expect that this situation can be improved by integrating ideas related to the transpension type~\cite{DBLP:journals/lmcs/NuytsD24,trascwod-hott-uf} into the current system.
Lastly, using $\sfswap$, we can follow Pitts et al.\ \cite{DBLP:journals/entcs/PittsMD15} in defining non-binding abstraction $\angles{x}- = \lambda\,a\,y.\,\sfswap\,x\,y\,a : A\,x \to (y : \aint) \multimap A\,y$, where $A$ can depend only on variables fresh for $x$.

\subparagraph{Locally scoped names}
Locally scoped names~\cite{DBLP:conf/popl/Odersky94,DBLP:conf/popl/Pitts10} are available in FreshMLTT, by the following rule on the left, and allow us to spawn a name from nowhere, provided that we use it to form a term that is fresh for it:
\begin{center}
	\AxiomC{$\Gamma, x : \aint \vdash T\,\sftype$}
	\noLine
	\UnaryInfC{$\Gamma, x : \aint \vdash t : T$}
	\noLine
	\UnaryInfC{$x$ is fresh for $t : T$}
	\UnaryInfC{$\Gamma \vdash \nu\,x.\,t : \nu\,x.\,T$}
	\DisplayProof
	\quad
	\begin{tabular}{l}
		$\Gamma, x : \aint \vdash \nu\,x.\,t = t : T$ \\
		Added: $\nu\,x.\,t = t$ if $x$ is not free in $t$.
	\end{tabular}
	\quad
	\AxiomC{$\Gamma \vdash S\,\sftype$}
	\noLine
	\UnaryInfC{$\Gamma, x : \aint \vdash t : S$}
	\noLine
	\UnaryInfC{$x$ is fresh for $t : S$}
	\UnaryInfC{$\Gamma \vdash \nu\,x.\,t : S$}
	\DisplayProof
\end{center}
It is used~\cite{nominal-sets-dtt-slides} in FreshMLTT to define e.g.\
\[
	\lambda\,c'.\,\nu\,x.\,\mathsf{case}\,(c'\,x)\,
	\left\{
		\begin{array}{l c l}
			\mathsf{inl}\,a \mapsto \mathsf{inl}\,\angles x a \\
			\mathsf{inr}\,b \mapsto \mathsf{inr}\,\angles x b
		\end{array}
	\right.	
	: (\aint \multimap A + B) \to (\aint \multimap A) + (\aint \multimap B)
\]
It has a computation rule which says that as soon as $x$ comes into scope again, we can drop the $\nu$-binder. Combined with $\alpha$-renaming, this means $\nu\,x.$ says `let $x$ be any name we have in scope, or a fresh one, it doesn't matter'. In particular, $\nu\,x.-$ is idempotent.
Before translating to nullary PTT, we add another equation rule, which says that we can drop $\nu\,x.-$ if $x$ is not used freely at all (in the example above, $x$ is used freely but freshly).
This way, the FreshMLTT $\nu$-rule above becomes equivalent to the one to its right. Indeed, the functions $T \mapsto \nu\,x.\,T$ and $S \mapsto S$ now constitute an isomorphism between types that are and are not dependent on $x : \aint$.
The advantage of the rule on the right is that it is not self-dependent.
In nullary PTT, we express freshness using $\sfGel$, suggesting the following adapted rules:
\begin{center}
	\AxiomC{$\Gamma, x : \aint \vdash t : \sfGel\,S\,x$}
	\UnaryInfC{$\Gamma \vdash \nu\,x.\,t : S$}
	\DisplayProof
	\qquad
	\begin{tabular}{l}
		$\Gamma, x : \aint \vdash \sfgel\,(\nu\,x.\,t)\,x = t : \sfGel\,S\,x$ \\
		$\nu\,x.\,\sfgel\,t\,x = t$ (where $x$ cannot be free in $t$ by \textsc{GelI}).
	\end{tabular}
\end{center}
In this formulation, it is now clear that $\nu x\ldotp t$ can be implemented as $\sfung~(\lambda x \ldotp t)$, while the two computation rules follow from \textsc{Gel}$\eta$ and \textsc{Gel}$\beta$.

\section{Nominal Pattern Matching}\label{sec:case-studies}

In this section we provide concrete examples of functions $D \to E$ defined by
recursion on a nominal data type $D$,
within PNTT, explained in \cref{sec:nullary-ptt}.

\subsection{Patterns That Bind}\label{sec:pi-calc-example}

Some nominal frameworks with existential name-abstraction
types~\cite{DBLP:conf/icfp/ShinwellPG03} provide
a convenient user interface to define functions $f:D \to E$ out
of a nominal data type. The user
can define $f$ by matching on patterns that
bind names (\EG{} $\mathsf{eqabs}$ in \cref{sec:intro}).
We explain how this feature is indirectly recovered in PNTT.

The following nominal data
type~\cite{DBLP:journals/corr/abs-0809-3960} is the nominal syntax of the $\pi$-calculus~\cite{DBLP:journals/iandc/MilnerPW92a},
a formal language
whose expressions represent
concurrently communicating processes.
The constructors stand for: terminate, silent computation step,
parallelism, non-determinism, channel allocation,
receiving\footnote{Binders
have arguments $\aint \multimap {-}$ and not $\nm \to {-}$ since the latter could lead to exotic process terms checking the bound name for equality to other names.}, and sending.
The arguments of (output) type $\mathsf{Proc}$ can be thought of as continuations, e.g.\ $\mathsf{nu}$ prepends a channel allocation instruction.
\begin{center}
	\begin{minipage}{.46\textwidth}
		\begin{lstlisting}[mathescape]
data Proc : $\cU$ where
  nil : Proc
  $\tau$pre : Proc $\to$ Proc
  par, sum : Proc $\to$ Proc $\to$ Proc\end{lstlisting}
	\end{minipage}
	\hspace{\stretch{1}}
	\begin{minipage}{0.5\textwidth}
		\begin{lstlisting}[mathescape]
  nu : ($\aint$ $\multimap$ Proc) $\to$ Proc
  inp : $\nm$ $\to$ ($\aint$ $\multimap$ Proc) $\to$ Proc
  out : $\nm$ $\to$ $\nm$ $\to$ Proc $\to$ Proc
		\end{lstlisting}
	\end{minipage}
	{\,}
\end{center}

An example of a process is
$\mathsf{par} \, (\mathsf{out}\,a\,b\,q) \allowbreak \, (\mathsf{inp} \, a\, (\lambda(x:\aint). \, p'x))$.
The first argument of $\mathsf{par}$ emits a name $b$ on channel $a$
and continues with $q$.
Simultaneously, the second argument waits for a name $x$
on channel $a$ and continues with $p'\,x$.
The expectation is that such an expression should reduce
to $\mathsf{par}\,q\,(p'\{b/x\})$ where $p'\{b/x\}$ replaces
occurrences of $(x:\aint)$ in the body of $p'$ by the name $b:\nm$.
Note that the application $p'\,b$ does not typecheck and that this substitution operation called $\mathsf{nsub}$ below
must be defined recursively, as done in~\cite{DBLP:journals/corr/abs-0809-3960}.
The following is an informal definition of $\mathsf{nsub}$ where some patterns bind (bridge) variables.

\begin{lstlisting}[mathescape]
nsub : $\nm$ $\to$ ($\aint$ $\multimap$ Proc) $\to$ Proc
nsub b ($\lambda$x. par (u' x) (v' x)) = par (nsub b u') (nsub b v')
$\ldots$ --nil, $\tau$pre, sum similar
nsub b ($\lambda$x. (nu (q' x))) = nu ($\lambda$y. nsub b ($\lambda$x. q' x y))
nsub b ($\lambda$x. inp a  (q' x)) = inp a ($\lambda$y. nsub b ($\lambda$x. q' x y)) --(0)
nsub b ($\lambda$x. inp ($\cart$ x)  (q' x)) = inp b ($\lambda$y. nsub b ($\lambda$x. q' x y)) --(1)
$\ldots$
\end{lstlisting}
Note how patterns (0) and (1) match on a binding pattern of the form $\lambda x.\,\mathsf{inp}~m~(q'~x)$, and cover the case where $m$ is different, resp.\ equal to the variable being substituted.
The informal $\mathsf{nsub}$ function reduces accordingly.

More formally in our system we define $\mathsf{nsub}$ by using the SAP
at $\mathsf{Proc}$, so $\mathsf{nsub}\,b$ is the following composition
$(\aint \multimap \mathsf{Proc}) \xrightarrow{\:\:\simeq\:\:} \mathsf{AProc} \xrightarrow{\qquad} \mathsf{Proc}$.
The SAP at $\mathsf{Proc}$ asserts that $(\aint \multimap \mathsf{Proc})$
is equivalent to $\mathsf{AProc}$, the nullary translation of $\mathsf{Proc}$
(its constructors use a suggestive $[-]_0$ notation):
\begin{center}
	\begin{minipage}{.39\textwidth}
		\begin{lstlisting}[mathescape]
data AProc : $\cU$ where
  [nil]$_0$ : AProc
  [$\tau$pre]$_0$ : AProc $\to$ AProc
  [par]$_0$ , [sum]$_0$ : AProc $\to$ AProc $\to$ AProc\end{lstlisting}
	\end{minipage}
	\hspace{\stretch{1}}
	\begin{minipage}{0.60\textwidth}
		\begin{lstlisting}[mathescape]
  [nu]$_0$ : ($\aint$ $\multimap$ AProc) $\to$ AProc
  [inp]$_0$ : $1 + \nm$ $\to$ ($\aint$ $\multimap$ AProc) $\to$ AProc
  [out]$_0$ : $1+\nm$ $\to$ $1+\nm$ $\to$ AProc $\to$ AProc
		\end{lstlisting}
	\end{minipage}
	{\,}
\end{center}
We can then define $\mathsf{nsub}' : \nm \to \mathsf{AProc} \to \mathsf{Proc}$ by induction on the second argument.
Indeed the informal clauses (0), (1) can be translated into formal ones using
the $[\mathsf{inp}]_0$ constructor. More precisely, clause (1) uses
the function $[\mathsf{inp}]_0\,(\mathsf{inl}\,\mathsf{tt})$ of type $(\aint \multimap \mathsf{AProc}) \to \allowbreak\mathsf{AProc}$
and clause (0) uses the function $\lambda(n:\nm).\,[\mathsf{inp}]_0\,(\mathsf{inr}\,n)$
of type $\nm \to \allowbreak (\aint \multimap \mathsf{AProc}) \to \allowbreak\mathsf{AProc}$.

\subsection{A HOAS Example}\label{sec:hoas-example}
In the example sketched below, we connect the nominal syntax of the untyped lambda calculus (ULC) to a nominal, higher-order abstract syntax (HOAS) representation.
The example is performed within PNTT plus a single (standard) binary parametricity axiom.
Our proof is a port to PNTT of an existing argument by Atkey~\cite{DBLP:conf/tlca/Atkey09}, which uses
external \emph{Kripke} binary parametricity to connect the De Bruijn (non-nominal)
syntax of ULC to a (non-nominal) HOAS representation.

This example serves two purposes.
The first is to illustrate, by porting a complex argument from one setting to the other, how the combination of nullary (for nominal reasoning) and binary parametricity allows for a reasoning style similar to that of Kripke binary parametricity, and moreover with closely related denotational semantics.
This topic is elaborated further at the end of this section.

Secondly, the example illustrates that more often than not, proving the correctness of a function $f : D \to E$ defined
by recursion out of a nominal data type $D$ requires computing the parametricity translation of $f$.
Indeed, for $\mathsf{bind} : (\aint \multimap D) \to D$ a binder (constructor) of $D$,
the $\beta$-rule of $D$ says that
the redex $f (\mathsf{bind}\,d')$
computes to a term involving $\aint \multimap f$, the action of $f$ on bridges (see \cref{sec:struct-abstr-princ}).
So proving $f$ correct often requires characterizing $\aint \multimap f$ up to propositional equality.
This characterization is the SAP for $f$ (also in \cref{sec:struct-abstr-princ})
$[f]_0 \equiv_{[D]_0 \to [E]_0} [f]_0^{\mathcal{O}}$, where
$[f]_0$ is the recursive translation of $f$ and $[f]_0^{\mathcal{O}} := \mathsf{SAP}^{-1}_E \circ (\aint \multimap f) \circ \mathsf{SAP}_D$
the observational parametricity of $f$.

SAP proofs for composite terms like $f$ are unwieldy because one has
to show why the SAP instances for the primitives in $f$ compose to the global one
and this involves manually reproducing the structure of the term $f$ at the level of SAP proofs.
For this reason some instances of the SAP for terms are assumed to hold below, without proof
(see paragraph ``Roundtrip at $\Ltm_j$'').
The unwieldiness of interval-based PTTs is a good selling point for \emph{observational} parametric
type theories (see \cref{sec:related-work}), where
the SAP is not needed since a first-class recursive translation attempts to replace $\aint \multimap {-}$ altogether.

As a side note regarding the above function $f : D \to E$,
its domain $D$ could feature a binder of type $(\aint \multimap \ldots \multimap \aint \multimap D) \to D$
binding $m$ fresh names. In that case proving the correctness of $f$ may involve
computing its $m$-fold iterated nullary parametricity, for the same reasons as above.
No such examples are investigated here.

Let us now get to the actual argument.
The nominal syntax of ULC is expressed as the following data type family $\Ltm$, parametrized
by a natural number $j:\mathtt{nat}$.
\begin{lstlisting}[mathescape]
data Ltm (j : nat) : $\cU$
  env : Fin j $\to$ Ltm j
  var : $\mathsf{Nm}$ $\to$ Ltm j
  app : Ltm j $\to$ Ltm j $\to$ Ltm j
  lam : ($\aint$ $\multimap$ Ltm j) $\to$ Ltm j
\end{lstlisting}
The type \lstinline!Fin j! is the finite type with $j$ elements $\{0, \ldots, j-1 \}$.
$\Ltm_j$ is a shorthand for $\Ltm\,j$. The types $\Ltm$ and $\Ltm_1$ of \cref{sec:intro}
are $\Ltm_0$ and $\Ltm_1$ respectively.

The corresponding HOAS representation, or encoding, is $\Henc_j : \cU$ defined below.
Since it uses a $\Pi$-type we say it is a $\Pi$-encoding.
\begin{lstlisting}[mathescape]
$\mathsf{HMod}$ (j : nat) = (H : $\cU$) $\times$ (Fin j $\to$ H) $\times$ ($\mathsf{Nm}$ $\to$ H) $\times$
  (H $\to$ H $\to$ H) $\times$ ((H $\to$ H) $\to$ H)

--projections
|_| : $\forall$ {j}. HMod j $\to$ $\cU$
|_| M = M .fst
envOf, varOf, appOf, hlamOf = $\ldots$ --other projections of $\mathsf{HMod}$

$\Henc$ (j : nat) = (M : HMod j) $\to$ |M|

$\NMod$ (j : nat) = (H : $\cU$) $\times$ (Fin j $\to$ H) $\times$ ($\mathsf{Nm}$ $\to$ H) $\times$
  (H $\to$ H $\to$ H) $\times$ (($\aint$ $\multimap$ H) $\to$ H)
\end{lstlisting}
The type of ``nominal models'' $\NMod_j$ is also defined as it will be useful later on.
The carrier function $|{-}|$ and other projections are defined similarly for nominal models.
Additionally we define explicit constructors $\mkhm_j, \mknm_j$ for $\HMod_j, \NMod_j$.
For instance $\mknm_j : (H : \cU) \to (\mathsf{Fin}\,j \to H) \to (\nm \to H)
\to (H \to H \to H) \to ((\aint \multimap H)\to H) \to \NMod_j$.

We will show that we can define maps between $\Ltm_j$ and the encoding $\Henc_j$ in both ways,
and sketch a conditional proof that they compose to the identity at $\Ltm_j$.
\begin{lstlisting}[mathescape]
$\sfubd$ : $\forall${j}. $\Henc_j$ $\to$ $\Ltm_j$
$\sftoh$ : $\forall${j}. $\Ltm_j$ $\to$ $\Henc_j$
rdt-Ltm : $\forall${j}. (t : $\Ltm_j$) $\to$ $\sfubd$($\sftoh_j$ t) $\equiv$ t
\end{lstlisting}

\subparagraph*{Unembedding}
We begin by defining the ``unembedding'' map
denoted by $\sfubd$,
which has a straightforward definition that does not involve
nominal recursion.
The name and the idea behind the
definition come from~\cite{DBLP:conf/tlca/Atkey09, DBLP:conf/haskell/AtkeyLY09}, where
Atkey et al.\ were interested in comparing (non-nominal) syntax
and HOAS $\Pi$-encodings using a strengthened form of binary parametricity called Kripke parametricity
(see end of section for a comparison with our system/model).
\begin{lstlisting}[mathescape]
$\sfubd$ {j} = $\lambda$(h : $\Henc_j$). h (LtmHMod j) where
    LtmHMod : $\forall$ j. $\HMod_j$
    LtmHMod j = $\mathsf{mkHM}_j$ $\Ltm_j$ env var app (hlamLtm j)
    hlamLtm : $\forall$ j.($\Ltm_j$ $\to$ $\Ltm_j$) $\to$ $\Ltm_j$
    hlamLtm j f = lam( $\lambda$(x:$\aint$). f(var ($\cart$ x)) )
\end{lstlisting}
So unembedding a HOAS term $h$ consists of applying $h$ at $\Ltm_j$.
This is possible thanks to the fact that $\Ltm_j$ can be equipped with a higher-order operation $\mathsf{hlamLtm}_j$.

\subparagraph*{Defining the map into HOAS}
The other map $\sftoh_j$ is defined by nominal recursion.
In other words it is defined using the eliminator of $\Ltm_j$.
We write the (uncurried) non-dependent eliminator as $\mathsf{rec}_j$.
Its type expresses that $\Ltm_j$ is the initial nominal model.
\begin{lstlisting}[mathescape]
$\mathsf{rec}_j$ : (N :  $\mathsf{NMod}_j$) $\to$ $\Ltm_j$ $\to$ |N|
\end{lstlisting}

Hence in order to define $\sftoh_j$ we need to turn its codomain $\Henc_j$ into a nominal model.
For fields in $\NMod_j$ that are not binders this is straightforward.
\begin{center}
	\begin{minipage}{.40\textwidth}
\begin{lstlisting}[mathescape]
envH : Fin j $\to$ $\Henc_j$
varH : $\mathsf{Nm}$ $\to$ $\Henc_j$
appH : $\Henc_j\to\Henc_j\to$ $\Henc_j$
\end{lstlisting}
	\end{minipage}
	\hspace{\stretch{1}}
	\begin{minipage}{.56\textwidth}
\begin{lstlisting}[mathescape]
envH k = $\lambda$ (M:$\HMod_j$). envOf M k
varH n = $\lambda$ (M:$\HMod_j$). varOf M n
appH u v = $\lambda$ M. appOf M (u M) (v M) 
\end{lstlisting}
	\end{minipage}
\end{center}
Additionally we need to provide a function $\mathsf{lamH} : (\aint \multimap \Henc_j) \to \Henc_j$.
This is done by using the SAP at $\Henc_j$, i.e.\ the following characterization of $(\aint \multimap \Henc_j)$.
\begin{theorem}\label{thm:hbump}
We have
$\sfeb_j : (\aint \multimap \Henc_j) \simeq \Henc_{j+1}$,
$\sfmb_j : (\aint \multimap \HMod_j) \simeq \HMod_{j+1}$,
$\sfnb_j : (\aint \multimap \NMod_j) \simeq \NMod_{j+1}$ and
$\sflb_j : (\aint \multimap \Ltm_j) \simeq \Ltm_{j+1}$.
\end{theorem}
Intuitively these results hold because $\aint \multimap {-}$ preserves all primitive type formers except $\nm$,
which gets translated to a sum $1 + \nm$.
So translating each of the aforementioned types boils
down to replacing $\nm$ with $1+\nm$, which refactors to incrementing $j$.
\begin{proof}
We only prove $\sfmb_j$. The proofs of $\sfeb_j$ and $\sfnb_j$ are similar, and
$\sflb_j$ is proved using an encode-decode argument~\cite{DBLP:journals/lmcs/CavalloH21, DBLP:journals/pacmpl/MuylderND24}
similar to the proof of $\mathsf{SAP}_\nm$.
We sometimes omit types for $\Sigma$ and $\multimap$, e.g.\
we write $x \multimap T$ as shorthand for $(x:\aint) \multimap T$.
\begin{align*}
x \multimap \HMod_j &\simeq (H' : x \multimap \cU) \times (x \multimap [(\mathsf{Fin}\,j \to H'x) \times \ldots]   )                  &\mathsf{SAP}_\Sigma\\
                    &\simeq (H' : x \multimap \cU) \times (x \multimap (\mathsf{Fin}\,j \to H'x)) \times (x \multimap [\ldots])       &\mathsf{SAP}_\Sigma\\
                    &\simeq (H' : x \multimap \cU) \times (\mathsf{env}' : \mathsf{Fin}\,j \to (x \multimap H'x)) \times (x \multimap [\ldots]                ) &\mathsf{SAP}_{\mathsf{Fin}\,j, \to}
\end{align*}
Since $\mathsf{Fin}\,j$ is a non-nominal data type its SAP instance asserts
$\mathsf{Fin}\,j \simeq (x\multimap \mathsf{Fin}\,j)$, i.e.\ the only bridges in $\mathsf{Fin}\,j$
are reflexive bridges~\cite{DBLP:journals/lmcs/CavalloH21, DBLP:journals/pacmpl/MuylderND24}. Moving on,
\begin{align*}
                   &\simeq H' \times \mathsf{env'} \times (x \multimap [(\nm \to H'x) \times \ldots])                                  &\\
                   &\simeq H' \times \mathsf{env'} \times (x \multimap (\nm \to H'x)) \times (x \multimap [\ldots])                   &\mathsf{SAP}_\Sigma\\
                   &\simeq H' \times \mathsf{env'} \times ((x \multimap \nm) \to (x \multimap H'x)) \times (x \multimap [\ldots])      &\mathsf{SAP}_\to\\
                   &\simeq H' \times \mathsf{env'} \times (\mathsf{foo} : (1 + \nm) \to (x \multimap H'x)) \times (x \multimap [\ldots]) &\mathsf{SAP}_\nm\\
                   &\simeq H' \times \mathsf{env'}_+ \times (\mathsf{var}' : \nm \to x \multimap H'x) \times (x \multimap [\ldots])             &
\end{align*}
where $\mathsf{env'}_+ : \mathsf{Fin}\,(j+1) \to (x \multimap H'x)$ is defined as $\mathsf{env'}_+\,0 = \mathsf{foo}\,(\mathsf{inl}\,\mathsf{tt})$
and $\mathsf{env'}_+\,(k+1) = \mathsf{env'} k$.
The next two types in $x \multimap [\ldots]$ are computed using the SAP at $\to$.
Thus so far we have shown
that $\aint \multimap \HMod_j$ is equivalent to
$(H' : x \multimap \cU) \times \mathsf{env'}_+ \times \mathsf{var}' \times (\mathsf{app}' : (x \multimap H'x) \to (x \multimap H'x) \to (x \multimap H'x))
\times  (\mathsf{hlam}' : ((x \multimap H'x) \to (x \multimap H'x)) \to (x \multimap H'x))$.
Now since the SAP at $\cU$ is  $\sfGel : \cU \overset{\simeq}{\to} (\aint \multimap \cU)$,
we can do a change of variable in the above translation of $\HMod_j$, i.e.\
use a variable $(K : \cU)$ instead of $(H' : x \multimap \cU)$ at the cost of
replacing occurrences of $H'$ by $\sfGel\,K$.
Pleasantly, occurrences of $(x \multimap H'x)$ become $(x \multimap \sfGel\,K\,x)$, which is equivalent to $K$
by the SAP for $\sfGel$ types.
Hence $\aint \multimap \HMod_j \simeq \HMod_{j+1}$.
\end{proof}

Next, we can define the desired operation $\mathsf{lamH} : (\aint \multimap \Henc_j) \to \Henc_j$
as $\mathsf{lamH}\,h' = \lambda(M:\HMod_j).\,\mathsf{lamOf}\,M\,( \,\lambda (m:|M|).\,(\sfeb_j\,h')\,(m :: M))$
where $(m :: M)$ is the $\HMod_{j+1}$ obtained out of $M : \HMod_{j}$ by pushing
$m$ onto the list $\mathsf{envOf} M$. In other words,
$\mathsf{envOf}\,(m::M)\,0 = m$ and $\mathsf{envOf}\,(m::M)\,(k+1) = \mathsf{envOf}\,M\,k$.
All in all we proved that $\Henc_j$ is a nominal model.
Since $\Ltm_j$ is the initial one we obtain the desired map $\sftoh_j : \Ltm_j \to \Henc_j$.
\begin{lstlisting}[mathescape]
$\sftoh_j$ = $\mathsf{rec}_j$ ($\mknm_j$ Henc$_j$ envH varH appH
  ($\lambda$ h' M. lamOf M ($\lambda$(m:|M|). $\sfeb_j$ h' (m::M))))
\end{lstlisting}

\subparagraph*{Roundtrip at $\Ltm_j$}
We sketch a conditional proof that $\forall j\,(t:\Ltm_j).\,t \equiv \sfubd_j\,(\sftoh_j t)$, by (nominal) induction,
\IE{} using the dependent eliminator of $\Ltm_j$.
Our proof is conditional since it relies on two axioms.\
(1) We use a specific axiom called $\mathsf{binx}$ that is an instance of standard binary parametricity.
It is natural to rely on such an axiom here since our proofs attempt to emulate Kripke \emph{binary} parametricity
(the Kripke aspect is in fact emulated by nullary parametricity).
An expected model for this standard binary parametricity statement is
$\mathsf{Psh}(\boxslash_2 \times \Box_0 \times \Box_2)$
where $\Box_2$ is the original binary affine cube category used by Cavallo and Harper (CH)~\cite{DBLP:journals/lmcs/CavalloH21} to model binary parametricity, see  \cref{sec:semantics-soundness}.
In fact, we expect $\mathsf{binx}$ to be provable in PNTT extended with CH internal binary parametricity operators.\ 
(2) Some steps require characterizing the observational parametricity $[-]_0^{\mathcal{O}}$ (defined in \cref{sec:struct-abstr-princ})
of certain terms. These steps are written with a \colorbox{gray!25}{gray background} below and
are \emph{assumed}.
As mentioned above characterizing the observational parametricity $[f]_0^{\mathcal{O}}$ of a complex term $f$ is
unwieldy because such characterizations must prove that $[-]_0^{\mathcal{O}}$ commutes with the
primitives used in $f$ and reproduce by hand the structure of the term $f$.
We currently have a partial paper proof of these steps. In the future
we would like to implement our theory (or maybe an observational version of it)
to fully validate this example.

The proofs for constructors other than $\mathsf{lam}_j$ are easy.
For $t = \mathsf{lam}_j (g:\aint \multimap \Ltm_j)$ we
must prove that if the induction hypothesis holds
$(g^\bullet : (z:\aint) \multimap (g\,z \equiv \sfubd_j\,(\sftoh_j (g\,z))))$ then
$\mathsf{lam}_j\,g \equiv \sfubd_j\,(\sftoh_j\,(\mathsf{lam}_j \,g))$.
Let $g^\bullet$ be in context and let us compute the right-hand side $\mathit{RHS}$ of the latter equation.
We use the $\$$ application operator found e.g.\ in Haskell. This operator is defined as
function application but  associates to the right, improving clarity.
\begin{align*}
\hspace{-1em}\mathit{RHS}
&\equiv \sfubd_j \, \$\, \lambda M.\,\mathsf{lamOf}\,M \,\$\, \lambda m.\, (\sfeb_j\circ (\aint \multimap \sftoh_j)) \, g\,(m::M) &(\text{def. }\mathsf{toh}_j)\\
&\equiv \mathsf{hlamLtm}_j \,\$\, \lambda m.\, (\sfeb_j\circ (\aint \multimap \sftoh_j)) \, g\,(m::\mathsf{LtmHMod}_j) &(\text{def. }\mathsf{ubd}_j)\\
&\equiv \mathsf{hlamLtm}_j \,\$\, \lambda m.\, (\sftoh_{j+1}\circ \sflb_j) \, g\,(m::\mathsf{LtmHMod}_j) &\colorbox{gray!25}{$([\sftoh_j]_0^{\mathcal{O}}\equiv \sftoh_{j+1})$}\\ 
&\equiv \mathsf{lam}_j \, \$\,\lambda (x:\aint).\,(\sftoh_{j+1}\circ \sflb_j) \, g\,(\mathsf{var}(\cart\,x)::\mathsf{LtmHMod}_j) &(\text{def. }\mathsf{hlamLtm}_j)\\
&\equiv \mathsf{lam}_j \, \$\,\lambda (x:\aint).\,(\sftoh_{j+1}\circ \sflb_j) \, g\: \$ & \\
&                        \qquad\qquad \mkhm_{j+1} \,\Ltm_j \,(\mathsf{var}(\cart\,x)::\mathsf{env}_j)\,\mathsf{var}_j\,\mathsf{app}_j\,\mathsf{hlamLtm}_j &(\text{def. }::)
\end{align*}
The latter model is of type $\HMod_{j+1}$ and we observe that it looks similar
to $\mathsf{LtmHMod}_{j+1} = \mkhm_{j+1} \,\Ltm_{j+1}\allowbreak \,\mathsf{env}_{j+1}\,\mathsf{var}_{j+1} \allowbreak \,\mathsf{app}_{j+1} \allowbreak \,\mathsf{hlamLtm}_{j+1} : \HMod_{j+1}$.
More formally, for $(x:\aint)$ in context, the graph of
$\mathsf{plug}_x : \Ltm_{j+1} \to \Ltm_j : u \mapsto \sflb_j^{-1}\,u\,x$
turns out be a structure-preserving relation
between the two models $\mathsf{LtmHMod}_{j+1},\,\allowbreak(\mathsf{var}(\cart\,x)::\mathsf{LtmHMod}_j): \HMod_{j+1}$.
This suggests using the binary parametricity of the
dependent function $\mathsf{inEnc}_g := (\sftoh_{j+1}\circ \sflb_j) \, g\allowbreak :(M:\HMod_{j+1}) \to M$ which appears in our computed $\mathit{RHS}$.
The binary parametricity of $\mathsf{inEnc}_g$ is
$[\mathsf{inEnc}_g]_2 : \forall (M_0\,M_1 : \HMod_{j+1})(R : [\HMod_{j+1}]_2 \,M_0\,M_1).\,\allowbreak|R|(\mathsf{inEnc}_g\,M_0)(\mathsf{inEnc}_g\, M_1)$,
where $[\HMod_{j+1}]_2 \,M_0\,M_1$ is defined to be the type of structure-preserving relations
between $M_0$, $M_1$, and $|R| : |M_0| \to |M_1| \to \cU$ extracts the carrier of $R$.
The specific binary parametricity axiom we use
is $\mathsf{binx} := [\mathsf{inEnc}_g]_2 \, \mathsf{LtmHMod}_{j+1}\,\allowbreak(\mathsf{var}(\cart\,x)::\mathsf{LtmHMod}_j) \,\allowbreak(\mathsf{Graph}(\mathsf{plug}_x))$,
\IE{} the dependent function $\mathsf{inEnc}_g$ commutes with the function $\mathsf{plug}_x$. We then have:
\begin{align*}
\mathit{RHS}&\equiv\mathsf{lam}_j \, \$\,\lambda (x:\aint).\,(\sftoh_{j+1}\circ \sflb_j) \, g\,(\mathsf{var}(\cart\,x)::\mathsf{LtmHMod}_j)&\\
&\equiv \mathsf{lam}_j \, \$\,\lambda (x:\aint).\,\mathsf{plug}_x\,\$\,(\sftoh_{j+1}\circ \sflb_j) \: g\:\mathsf{LtmHMod}_{j+1}&(\mathsf{binx})\\
&\equiv \mathsf{lam}_j \, \$\,\lambda (x:\aint).\,\mathsf{plug}_x\,\$\,(\sfubd_{j+1}\circ\sftoh_{j+1})(\sflb_j\,g)&(\text{def. }\mathsf{ubd}_{j+1})\\
&\equiv \mathsf{lam}_j \, \$\,\lambda (x:\aint).\,(\sflb_j^{-1}\$\,(\sfubd_{j+1}\circ\sftoh_{j+1})(\sflb_j\,g))\,x&(\text{def. }\mathsf{plug}_x)
\end{align*}
Now it remains to show that the latter $\equiv \mathsf{lam}_j \, \$ \,\lambda (x:\aint).\, g\,x$, \IE{} that
the following equality holds $\mathsf{eqNextWorld}: \sflb_j\,g \equiv (\sfubd_{j+1}\circ\sftoh_{j+1})(\sflb_j\,g)$.
Let us write $r_j := \sfubd_j \circ \sftoh_j$.\
(1) The induction hypothesis tells us $g^\bullet : (z:\aint) \multimap (g\,z \equiv r_j (g\,z))$
so by the SAP at $\equiv$ we get $g \equiv \lambda(z:\aint).\,r_j (g z)$
and by applying $\sflb_j$ we get $\sflb_j\,g \equiv \sflb_j (\lambda(z:\aint).\,r_j (g z))$.\
(2) The observational parametricity of $r_j$ is such that
\colorbox{gray!25}{$[r_j]_0^{\mathcal{O}} \equiv [\sfubd_j \circ \sftoh_j]_0^{\mathcal{O}} \equiv [\sfubd_j]_0^{\mathcal{O}} \circ [\sftoh_j]_0^{\mathcal{O}} \equiv \sfubd_{j+1} \circ \sftoh_{j+1} = r_{j+1}$}. 
 The function $r_j$ has type $\Ltm_j \to \Ltm_j$
so $[r_j]_0^{\mathcal{O}} \equiv r_{j+1}$ entails (by def.\ of $[r_j]_0^{\mathcal{O}}$, see \cref{sec:struct-abstr-princ})
$r_{j+1} \circ \sflb_j \equiv \sflb_j \circ (\aint \multimap r_j)$.
Applying both sides to $g :\aint \multimap \Ltm_j$ we
get $r_{j+1} ( \sflb_j \,g) \equiv \sflb_j(\lambda (z:\aint).\,r_j(g \,z))$.\
(3) Using step (1), (2) and transitivity of $\equiv$ we get
$\sflb_j g \equiv r_{j+1} ( \sflb_j \,g) = \sfubd_{j+1}\,(\sftoh_{j+1}\,(\sflb_j\,g)) $.
Hence $\mathsf{eqNextWorld}$ holds and this concludes our proof.

\subparagraph*{Synthetic Kripke parametricity}
Kripke parametricity is a strengthened form of binary parametricity
that was introduced in a denotational model built by Atkey~\cite{DBLP:conf/tlca/Atkey09}
in order to prove an adequacy result for HOAS:
the De Bruijn syntax of ULC
is equivalent to the HOAS $\Pi$-encoding $\forall A. (A \to A \to A) \to ((A \to A) \to A) \to A$.
One way to explain Kripke binary parametricity is to compare it to standard binary parametricity.
The latter asserts that polymorphic functions $f : (A : \cU) \to T(A)$
map relations $R$ to proofs of relatedness over $R$ (this
is explained for $T(A) = A \to A$ in \cref{sec:intro}).
By contrast, briefly speaking, the \emph{Kripke} binary parametricity of a polymorphic function $f : (A : \cU) \to T(A)$
is written $[f]_{\mathsf{K}2}$ and
asserts that for every (*) preorder $W$, $f$ maps any $W$-monotonic
family of relations $(w \mapsto R_w) : W \to A_0 \to A_1 \to \cU$
to a monotonic function/proof that for every $w:W$ the outputs $f\,A_0,f \, A_1$ are related over $R_w$
($\cU$ is implicitly ordered by logical implication).

We think that the example discussed above externalizes to
the fragment of Atkey's Kripke model where the $W$ preorder is systematically replaced by
$(\mathbb{N},\leq)$, natural numbers with the usual ordering.\footnote{A mismatch:
our model uses the category $\Box_0$ (\cref{sec:semantics-soundness}) while Atkey's uses \emph{preorders} like $(\mathbb{N},\leq)$.}
Indeed the example above is performed within PNTT with an additional binary parametricity axiom $\mathsf{binx}$.
The idea is then that the nullary parametricity aspect of our argument gets
translated to the monotonic quantification aspect of Atkey's Kripke model.
That is, $(0,2)$-ary parametricity emulates $(\mathbb{N},\leq)$-restricted Kripke $2$-ary parametricity.

Regarding unrestricted Kripke parametricity,
we observe that the quantification (*) on preorders is \emph{hard-coded} into $[f]_{\mathsf{K}2}$,
i.e.\ $[f]_{\mathsf{K}2}$ is not the mere meta-theoretical conjunction of its $W$-restricted Kripke parametricities.
In particular $[f]_{\mathsf{K}2}$ lives one universe level higher than $f$ which is peculiar.
Interestingly this hard-coding plays a crucial role in Atkey's proof of the roundtrip at $\forall A. (A \to A \to A) \to ((A \to A) \to A) \to A$.
Indeed, within Atkey's model, the proof introduces a variable $B:\cU$ and uses
$[f]_{\mathsf{K}2}$ at preorder $W:= \mathsf{List}\,B$ with prefix ordering, and
at type $A_0 := B$ (and some other type for $A_1$). So $B$ intervenes
both to determine the kind of parametricity being used $(W := \mathsf{List}\,B)$, and as
an input for this parametricity ($A_0 := B$).
We see this as a form of diagonalization, which we were not able to internalize.

\section{Related and Future Work}\label{sec:related-work}

We have already discussed the most closely related work in
internally parametric type theory~\cite{bernardy2015presheaf,DBLP:journals/lmcs/CavalloH21,DBLP:journals/pacmpl/MuylderND24} and in nominal type theory~\cite{DBLP:conf/icfp/ShinwellPG03,DBLP:phd/ethos/Schopp06,schopp2004dependent,DBLP:journals/corr/abs-1201-5240,DBLP:journals/entcs/PittsMD15}
in detail in Sections~\ref{sec:nullary-ptt} and~\ref{sec:nomin-prim-free} respectively.
Note that a subtle error in the construction of the model shown in~\cite{bernardy2015presheaf} was recently
spotted in~\cite{DBLP:journals/pacmpl/AltenkirchCKS24} and investigated in~\cite{reboullet:tel-05584055}.

Regarding the semantics of nominal frameworks,
PNTT is modelled (see \cref{sec:semantics-soundness}) in
$\mathsf{Psh}(\boxslash_2 \times \Box_0)$, \IE{} bicubical sets over the binary cartesian cube category $\boxslash_2$ and the nullary affine cube category $\Box_0$.
The latter is also the base category of the Schanuel sheaf topos~\cite[\S 6.3]{pitts-nominal-sets}, which is in turn equivalent to the category of nominal sets~\cite{pitts-nominal-sets}, used to model FreshMLTT~\cite{DBLP:journals/entcs/PittsMD15}.
FreshML~\cite{DBLP:conf/icfp/ShinwellPG03} is modelled in the Fraenkel-Mostowski model of set theory,
which inspired the development of nominal sets.
Sch\"opp and Stark's bunched system~\cite{schopp2004dependent} is modelled in a class of categories, including the Schanuel topos, that are locally cartesian closed and also equipped with a semicartesian closed structure.
Finally, \sysCheney{}~\cite{DBLP:journals/corr/abs-1201-5240} has a syntactic soundness proof.

$\Box_0$ is a category of cubes without diagonals,
so the internal language of its associated presheaf topos is among the first candidates
to get a transpension type~\cite{DBLP:journals/lmcs/NuytsD24} with workable typing rules~\cite{trascwod-hott-uf}.
Dual to the fact that universal and existential name quantification semantically coincide, so do freshness and transpension.
As such, the transpension type is already present as $\sfGel$ in the current system, but $\sfGel$'s typing rules are presently weaker: the rules \textsc{GelF} and \textsc{GelI} remove a part of the context, rather than quantifying it.
The relationship between $\sfGel$ and transpension is explained in more generality in~\cite{DBLP:journals/lmcs/NuytsD24}.

Regarding the implementation of PNTT,
given that the binary CH theory was implemented in~\cite{DBLP:journals/pacmpl/MuylderND24} on top of Cubical
Agda~\cite{vezzosi_cubical_2019},
implementing PNTT (essentially nullary CH + name induction) is feasible.
We think that such an implementation would be usable but not particularly scalable by default to perform
parametricity proofs.
The issue is that such proofs require the SAP, which cannot be proved as a single
internal theorem (see \cref{sec:hoas-example} and~\cite{DBLP:journals/pacmpl/MuylderND24}).
A practical solution could be to automate the process of building SAP proofs.
Alternatively, one could instead build PNTT on top of
\emph{observational}~\cite{DBLP:journals/pacmpl/AltenkirchCKS24} nullary PTT,
where the SAP is not needed since a first-class recursive translation attempts to replace the Bridge type.
The Narya proof assistant in progress~\cite{narya_nominal2025} implements observational $n$-ary PTT (and more).



\bibliography{references}

\end{document}